\documentclass[aps,pra,reprint,floatfix,superscriptaddress,longbibliography]{revtex4-2}

\usepackage[T1]{fontenc}
\usepackage[utf8]{inputenc}
\usepackage{amsmath,amssymb,amsfonts,mathtools,bm}
\usepackage{braket}
\usepackage{physics}
\usepackage{graphicx}
\usepackage{xcolor}
\usepackage{hyperref}
\usepackage{tikz}
\usetikzlibrary{arrows.meta,calc,positioning,decorations.pathreplacing}

\hypersetup{
    colorlinks=true,
    linkcolor=blue,
    citecolor=blue,
    urlcolor=blue
}

\begin{document}

\title{Bound-state-protected phase metrology for a quantum emitter in a Su--Schrieffer--Heeger bath}

\author{Sofia Evangelou}
\affiliation{School of Electrical and Computer Engineering, Technical University of Crete, Chania 73100, Greece}
\affiliation{Institute for Quantum Computing and Quantum Technologies, NCSR Demokritos, Greece}

\date{\today}

\begin{abstract}
We study local phase estimation for a single quantum emitter coupled to a bosonic Su--Schrieffer--Heeger (SSH) bath within a microscopic lattice model. Dimerization opens a central gap supporting an in-gap emitter-bath bound state, which suppresses complete relaxation of the emitter coherence. A Dyson-equation analysis yields the local bath Green's function, the in-gap bound-state condition, and the emitter residue controlling the retained phase information. The phase quantum Fisher information links gap formation, detuning, and emitter-bath coupling to the post-transient metrological response. At resonance, stronger coupling enhances transient hybridization but reduces the retained signal by lowering the emitter weight in the bound state. Away from resonance, late-time averages, retention times, and useful interrogation windows track how phase-information protection weakens as the emitter is tuned toward and beyond the band edge. A uniform-chain control shows that the retained signal disappears when the gap closes. In the bulk local-coupling geometry considered here, the response is insensitive to the sign of the dimerization, so the protocol probes spectral-gap physics and bound-state support rather than the SSH winding sector.
\end{abstract}

\maketitle

\section{Introduction}

Quantum metrology addresses the estimation of unknown parameters with a precision ultimately limited by quantum mechanics. Its central figure of merit is the quantum Fisher information (QFI), which sets the attainable sensitivity through the quantum Cram\'er--Rao bound and quantifies the local distinguishability of nearby quantum states \cite{BraunsteinCaves1994}. The general framework of quantum-enhanced estimation is now well established, both in broad reviews and in platform-oriented sensing settings \cite{Giovannetti2011,Degen2017,Pezze2018}. In idealized scenarios, the achievable precision is determined by the probe state, the encoding dynamics, and the final measurement. In realistic platforms, however, the probe is unavoidably coupled to an environment. Decoherence and relaxation then degrade the encoded information and often become the dominant limitation on precision. This has made noisy and open-system quantum metrology a central topic of study, including from the perspective of dissipative precision bounds and explicitly non-Markovian processes \cite{Alipour2014,Chin2012,Altherr2021}.

Structured reservoirs are relevant in this context because their spectral features can qualitatively reorganize metrological dynamics. Memory effects, incomplete decay, and emitter-reservoir bound states can slow the loss of information, generate revivals, and in favorable regimes preserve a finite QFI at long times \cite{Chin2012,Lu2010,Breuer2016}. In photonic band-gap media, the bound-state picture goes back to the dressed-atom analysis of John and Wang \cite{JohnWang1990}, while modern waveguide-QED realizations have made the same physics directly accessible in slow-light and photonic-crystal platforms \cite{Calajo2016,LiuHouck2017,Sundaresan2019}. Closely related light-matter phenomena have also been explored in other structured photonic settings, including photonic Weyl systems \cite{GarciaElcano2021} and nanophotonic platforms based on two-dimensional and topological materials \cite{ThanopulosGraphene2022,ThanopulosTIOL2022,ThanopulosTINano2023}. In such settings, non-Markovian spontaneous-emission dynamics, strong light-matter coupling, population trapping, and bound-state effects have been analyzed for emitters near graphene nanodisks \cite{ThanopulosGraphene2022} and topological-insulator nanospheres \cite{ThanopulosTIOL2022,ThanopulosTINano2023}. Non-Markovian radiative effects have also recently been observed directly in structured photonic lattices \cite{Vicencio2025}.

Within quantum metrology, suppression of spontaneous emission near or inside a spectral gap can protect phase-estimation precision and lead to long-time QFI trapping \cite{Berrada2015}. Related behavior has been reported in Lorentzian and nonperfect band-gap reservoirs, where the temporal profile of the QFI is strongly shaped by the bath spectrum and the resulting non-Markovian decay dynamics \cite{Berrada2013,Huang2018,Mirkin2020}. Exact and numerically controlled open-system analyses have further shown that metrological performance can be enhanced by non-Markovian memory and by going beyond weak-coupling or rotating-wave approximations \cite{Wu2020,Zhang2022,Wu2021}. Taken together, these works show that reservoir structure can substantially modify the temporal profile of the QFI in open-system parameter-estimation problems \cite{Chin2012,Berrada2013,Berrada2015,Bai2019,Altherr2021}.

It is therefore useful to formulate such metrological questions in microscopic platforms whose band structure and local light-matter coupling are explicit from the outset. This shifts the discussion from an externally prescribed spectral density to a concrete lattice Hamiltonian with well-defined microscopic control parameters. Related microscopic reservoir programs have been developed in structured-waveguide, circuit-QED, and lattice-emitter settings \cite{Calajo2016,LiuHouck2017,Sundaresan2019}, as well as in more unconventional lattice-waveguide geometries such as Hofstadter-ladder baths \cite{WangHofstadter2022}. In such settings one can ask, in a controlled way, how band gaps, detuning, coupling strength, and bound-state formation determine the retention of metrologically useful information.

The bosonic Su--Schrieffer--Heeger (SSH) lattice provides a suitable microscopic setting for this question \cite{Ozawa2019,Bello2019}. Its dimerized structure generates two bands separated by a tunable central gap, while local coupling to a quantum emitter gives access to bound states, anomalous radiative dynamics, vacancy-like dressed states, and other structured-bath effects that have been studied in waveguide-QED and topological-photonic settings \cite{Bello2019,Ozawa2019,Sheremet2023,Leonforte2021,Gao2024}. Experimental interfaces between emitters and topological photonic modes have been realized in photonic-crystal and topological-waveguide architectures \cite{Barik2018,Kim2021}, and related theoretical work has explored few-body and multi-excitation bound states in analogous topological waveguide-QED geometries \cite{Vega2021,Vega2023}. Recent work has further clarified the role of weak-coupling bound states in semi-infinite topological waveguides and, more generally, the emergence of Hermitian and non-Hermitian topology in photon-mediated settings \cite{Garmon2026,RoccatiNatComm2024}. These developments make the SSH environment a concrete lattice model in which to examine how structured-bath physics controls dissipative phase metrology.

At the same time, it is useful to distinguish between metrological signatures of structured-spectrum physics and those of topology in a stricter sense. More explicitly topology-oriented sensing protocols have recently been proposed in related SSH-waveguide settings, particularly in boundary-coupled geometries where paired topological bound states can be used as an effective sensing resource \cite{TopoSensor2025}. The present bulk single-emitter setup plays a different role: it isolates the metrological effects associated with gap formation, band-edge structure, and bound-state support, without directly probing the SSH winding sector.

In this work, we study local phase estimation for a single quantum emitter coupled to a bosonic SSH bath. Our goal is to determine how dimerization, detuning, and coupling strength reorganize the phase-QFI through gap formation, band-edge effects, and bound-state support. The problem remains microscopic throughout: the metrological analysis is carried out directly within the lattice Hamiltonian, rather than through a phenomenological master equation or an imposed spectral density.

The analysis gives three main results. First, the reduced single-emitter state is obtained in closed form and gives the exact phase-QFI $F_Q^{(\phi)}(t)=|u(t)|^2$, where $u(t)$ is the excited-state survival amplitude. Second, a Dyson-equation analysis of the single-excitation Green's function yields the closed-form local bath Green's function, the central in-gap pole equation, and the emitter residue $Z_{\mathrm{BS}}(\Delta)$. This residue provides the bound-state benchmark for the retained phase information in the parameter regimes considered here. It also explains the sign independence of the bulk protocol through the $d^2$ structure of the local Green's function and predicts the suppression of the retained signal as the in-gap bound state approaches a band edge and its emitter residue vanishes. Third, the numerical analysis shows how dimerization, detuning, and coupling strength reorganize both the late-time signal and the transient dynamics. At resonance, increasing $g$ reduces the retained late-time signal while strengthening the transient oscillatory hybridization. Away from resonance, the response is analyzed through late-time averages, retention times, and post-transient useful interrogation windows. These diagnostics track the loss of protection as the emitter is tuned toward and beyond the band edge. The detuning-resolved numerics further show that the finite-window late-time signal tracks the detuning dependence of the central in-gap bound-state residue closely. In the present bulk local-coupling geometry, the local phase-QFI therefore resolves gap physics and bound-state support but does not distinguish the SSH winding sector.

Taken together, these results provide a microscopic SSH-lattice setting for structured-environment phase metrology and give a physical picture of how band engineering, coupling, and bound-state support govern late-time phase-information retention.

The paper is organized as follows. In Sec.~\ref{sec:model}, we introduce the model and the phase-estimation protocol. In Sec.~\ref{sec:qfi}, we derive the reduced state of the emitter and obtain the exact phase-QFI in closed form. In Sec.~\ref{sec:results}, we analyze the resulting dynamics in terms of the SSH-bath parameters and discuss the roles of dimerization, detuning, coupling strength, and bound-state protection. Section~\ref{sec:conclusions} contains the conclusions of the article.

\section{Model and numerical implementation}
\label{sec:model}

\begin{figure}[tbp]
    \centering
    \includegraphics[width=0.98\columnwidth]{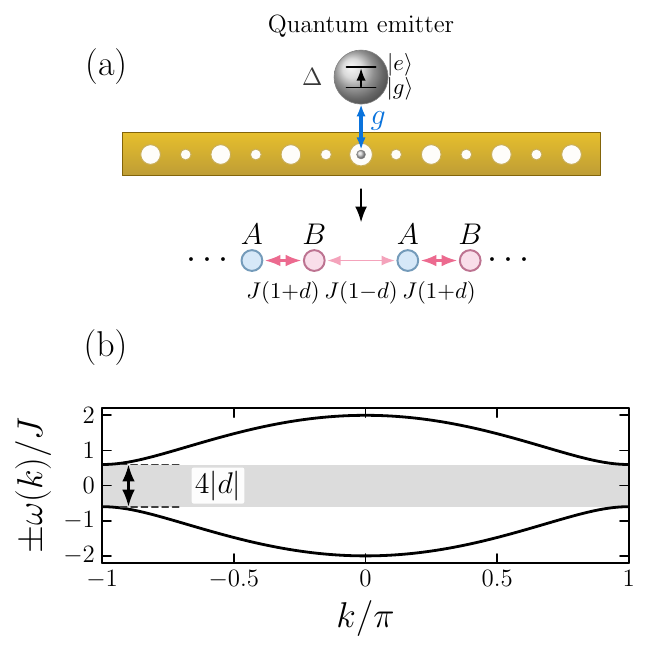}
    \caption{Model overview. (a) Single quantum emitter locally coupled with strength $g$ to a photonic structure that leads to a sublattice of a dimerized bosonic SSH chain with intracell and intercell hoppings $J_1=J(1+d)$ and $J_2=J(1-d)$. (b) SSH dispersion for a representative dimerization, shown together with the central band gap $4J|d|$.}
    \label{fig:model_schematic}
\end{figure}

We consider a two-level emitter with ground state $\ket{g}$ and excited state $\ket{e}$, locally coupled to a one-dimensional bosonic SSH bath \cite{Bello2019}. The environment contains two sublattices, $A$ and $B$, per unit cell and alternating nearest-neighbor hoppings. This dimerized lattice realizes a structured continuum with two bands separated by a central gap controlled by the dimerization parameter $d$. The physical setting is illustrated in Fig.~\ref{fig:model_schematic}.

The Hamiltonian is
\begin{equation}
H = H_{\mathrm S} + H_{\mathrm B} + H_{\mathrm{int}},
\label{eq:total_hamiltonian}
\end{equation}
with emitter term
\begin{equation}
H_{\mathrm S} = \Delta\,\sigma_{ee},
\label{eq:emitter_hamiltonian}
\end{equation}
bath term
\begin{equation}
H_{\mathrm B} = \sum_n \Big[ J(1+d)a_n^{\dagger}b_n + J(1-d)a_{n+1}^{\dagger}b_n + \mathrm{H.c.} \Big],
\label{eq:bath_hamiltonian}
\end{equation}
and local emitter-bath coupling
\begin{equation}
H_{\mathrm{int}} = g\left(\sigma_{eg}a_0 + \sigma_{ge}a_0^{\dagger}\right).
\label{eq:interaction_hamiltonian}
\end{equation}
Here $a_n$ and $b_n$ annihilate bosonic excitations on the two sublattices of unit cell $n$, $\sigma_{ee}=\ket{e}\!\bra{e}$, $\sigma_{eg}=\ket{e}\!\bra{g}$, and $\Delta$ is the emitter detuning measured from the middle of the gap. The bath dispersion is
\begin{equation}
\begin{aligned}
\omega_{\pm}(k)
&= \pm\sqrt{J_1^2 + J_2^2 + 2J_1J_2\cos k} \\
&= \pm 2J\sqrt{\cos^2(k/2)+d^2\sin^2(k/2)},
\end{aligned}
\label{eq:ssh_dispersion}
\end{equation}
with $J_1=J(1+d)$ and $J_2=J(1-d)$. The last form displays the gap structure explicitly: the band edges lie at $\omega=\pm 2J|d|$ (attained at $k=\pm\pi$), so the central gap has width $4J|d|$, while the outer edges lie at $\omega=\pm 2J$ (attained at $k=0$). In the present bulk geometry, these spectral scales provide the framework used below to analyze the metrological response.

Because the total excitation number is conserved, the dynamics generated by Eqs.~\eqref{eq:total_hamiltonian}--\eqref{eq:interaction_hamiltonian} closes in the single-excitation sector relevant to the present phase-estimation protocol. The total state can therefore be written as
\begin{equation}
\ket{\Psi(t)}
=
u(t)\ket{e}\ket{\mathrm{vac}}
+
\sum_{\lambda}\beta_\lambda(t)\ket{g}\ket{1_\lambda},
\label{eq:total_state_single_excitation}
\end{equation}
where $\ket{\mathrm{vac}}$ is the bath vacuum, $\ket{1_\lambda}$ denotes a single bath excitation, and $u(t)$ is the excited-state survival amplitude. Normalization gives
\begin{equation}
|u(t)|^2 + \sum_\lambda |\beta_\lambda(t)|^2 = 1.
\label{eq:normalization}
\end{equation}

To probe phase sensitivity, we take the emitter to be initially prepared in the balanced superposition
\begin{equation}
\ket{\psi_\phi(0)}
=
\frac{1}{\sqrt{2}}
\left(
\ket{g}+e^{i\phi}\ket{e}
\right),
\label{eq:initial_probe_state}
\end{equation}
with unknown phase $\phi$. This standard choice encodes the parameter in the ground-excited coherence and provides a natural local probe for dissipative phase estimation \cite{BraunsteinCaves1994,Degen2017,Pezze2018,Berrada2015,Berrada2013,Huang2018}. With the bath initially in the vacuum state, the total initial state is
\begin{equation}
\ket{\Psi_\phi(0)}
=
\frac{1}{\sqrt{2}}
\left(
\ket{g}+e^{i\phi}\ket{e}
\right)\otimes\ket{\mathrm{vac}}.
\label{eq:initial_total_state}
\end{equation}
At zero temperature, the ground component remains unchanged, while the excited component undergoes the nontrivial lattice-induced dynamics of Eq.~\eqref{eq:total_state_single_excitation}. The time-evolved state is therefore
\begin{equation}
\begin{aligned}
\ket{\Psi_\phi(t)}
&=
\frac{1}{\sqrt{2}}
\bigg[
\ket{g}\ket{\mathrm{vac}}
+
e^{i\phi}
\big(
u(t)\ket{e}\ket{\mathrm{vac}}\\
&+
\sum_\lambda \beta_\lambda(t)\ket{g}\ket{1_\lambda}
\big)
\bigg].
\label{eq:time_evolved_total_state}
\end{aligned}
\end{equation}

The numerical analysis is performed directly on finite open SSH chains centered on the emitter cell, with chain lengths chosen so that all displayed time windows remain below the first finite-size recurrences. For each parameter set, we construct the Hamiltonian in the single-excitation basis with the emitter coupled to the $A$ sublattice at $n=0$, and evaluate the survival amplitude
\begin{equation}
u(t) = \langle e|e^{-iHt}|e\rangle,
\label{eq:survival_amplitude}
\end{equation}
by one of two equivalent strategies, depending on the chain length used for convergence. For moderate chain sizes we obtain $u(t)$ from the full spectral decomposition of the sparse single-excitation Hamiltonian. For the larger chains, where full diagonalization is impractical, we first Lanczos-tridiagonalize $H$ starting from the initial emitter state $\ket{e}$ and then diagonalize the resulting tridiagonal representation. Because the dynamics is needed only for the projection onto $\ket{e}$, the Krylov subspace generated from this seed state gives $u(t)$ exactly up to the reported tolerance. In practice we use a Lanczos breakdown tolerance of $10^{-12}$, which is never triggered within the truncation dimensions stated in the captions. Each figure caption specifies which strategy is used. Both procedures yield the emitter dynamics directly from the microscopic lattice Hamiltonian, without introducing an external spectral density or a phenomenological master equation.

The chain sizes and, where applicable, Krylov dimensions are stated explicitly in each figure caption. They are chosen so that the time traces and late-time indicators reported in the manuscript are stable within the displayed windows. Representative convergence checks show that increasing the chain length further leaves the plotted pre-recurrence observables unchanged within plotting resolution, while enlarging the Krylov dimension, where used, produces changes smaller than the line width in representative gapped, near-edge, and continuum regimes. Strict finite-chain infinite-time averages converge more slowly because they are more sensitive to grouped quasi-degeneracies in the discrete spectrum. For this reason, the manuscript adopts finite-window late-time indicators as the primary off-resonant operational diagnostics. Throughout, we focus on the bulk configuration defined by local coupling to the $A$ sublattice at $n=0$, so that the reported trends isolate the effect of the SSH band structure and central gap rather than edge-state physics.

\section{Phase-QFI formalism}
\label{sec:qfi}

We now connect the microscopic emitter dynamics to the phase-estimation task. Tracing over the bath degrees of freedom gives the reduced emitter state
\begin{equation}
\rho_\phi(t)=\Tr_B\!\left[\ket{\Psi_\phi(t)}\bra{\Psi_\phi(t)}\right].
\label{eq:reduced_density_def}
\end{equation}
Using Eq.~\eqref{eq:normalization}, this becomes
\begin{equation}
\rho_\phi(t)=
\begin{pmatrix}
1-\dfrac{|u(t)|^2}{2} & \dfrac{u^*(t)e^{-i\phi}}{2}\\[8pt]
\dfrac{u(t)e^{i\phi}}{2} & \dfrac{|u(t)|^2}{2}
\end{pmatrix},
\label{eq:reduced_density_matrix}
\end{equation}
in the ordered basis $\{\ket{g},\ket{e}\}$. The phase parameter therefore remains encoded in the local emitter coherence, while the environment enters only through the time-dependent survival amplitude $u(t)$.

To evaluate the corresponding metrological sensitivity, we write the reduced state in Bloch form
\begin{equation}
\rho_\phi(t)=\frac{1}{2}\left[I+\mathbf{r}(t,\phi)\cdot\boldsymbol{\sigma}\right],
\label{eq:bloch_form}
\end{equation}
with components
\begin{align}
r_x(t,\phi)&=\Re\!\left[u(t)e^{i\phi}\right],\\
r_y(t,\phi)&=\Im\!\left[u(t)e^{i\phi}\right],\\
r_z(t)&=|u(t)|^2-1.
\end{align}
The squared Bloch-vector length is
\begin{equation}
|\mathbf{r}(t,\phi)|^2
=
|u(t)|^2+\left(|u(t)|^2-1\right)^2
=
1-|u(t)|^2+|u(t)|^4,
\label{eq:bloch_norm_2}
\end{equation}
which is independent of $\phi$. Therefore
\begin{equation}
\mathbf{r}(t,\phi)\cdot \partial_\phi \mathbf{r}(t,\phi)=0,
\label{eq:orthogonality_condition}
\end{equation}
and the one-parameter qubit QFI reduces to the standard Bloch-vector expression \cite{BraunsteinCaves1994}
\begin{equation}
F_Q^{(\phi)}(t)=\left|\partial_\phi \mathbf{r}(t,\phi)\right|^2.
\label{eq:qfi_bloch}
\end{equation}
Evaluating the derivative gives the exact phase-QFI of the present protocol,
\begin{equation}
F_Q^{(\phi)}(t)=|u(t)|^2.
\label{eq:main_qfi_result}
\end{equation}

Equation~\eqref{eq:main_qfi_result} fixes the phase-QFI in terms of the emitter survival probability. The bound is locally saturable by a single-qubit measurement. Since $\partial_\phi \mathbf{r}(t,\phi)$ lies in the equatorial plane of the Bloch sphere and is orthogonal to $\mathbf{r}(t,\phi)$ [Eq.~\eqref{eq:orthogonality_condition}], the optimal projective basis is the equatorial basis rotated by $\arg u(t)+\phi$. This should be understood as the locally optimal measurement around a prior estimate of $\phi$, or equivalently as the measurement implemented in an adaptive phase-estimation scheme. No ancilla-assisted protocol is required. In the rest of the paper, Eq.~\eqref{eq:main_qfi_result} provides a compact route to the central physical question: how the microscopic dimerized reservoir controls the persistence of phase sensitivity through gap formation, band-edge effects, and bound-state support \cite{Chin2012,Lu2010,Berrada2015,Berrada2013,Huang2018,Bello2019}.

\section{Results}
\label{sec:results}

We now analyze dissipative phase metrology in the single-emitter SSH setting, focusing on how the bath parameters reshape $F_Q^{(\phi)}(t)$ and thereby control late-time phase-information retention. In the present bulk geometry, the relevant mechanisms are set by the SSH spectral gap, the position of the emitter frequency relative to the band edges, the emitter-bath coupling strength, and the bound-state weight supported by the microscopic lattice spectrum. We first examine the resonant dependence on dimerization and coupling strength together with the bound-state origin of the retained late-time signal. We then use a Green's-function analysis to identify the central in-gap bound state and its emitter residue at arbitrary detuning, and finally use this bound-state benchmark to organize the off-resonant crossover, the associated operational diagnostics, and the detuning-resolved microscopic bound-state analysis.

\subsection{Resonant analysis and bound-state origin of the retained late-time signal}
\label{sec:results_resonant}

\begin{figure}[tbp]
    \centering
    \includegraphics[width=\columnwidth]{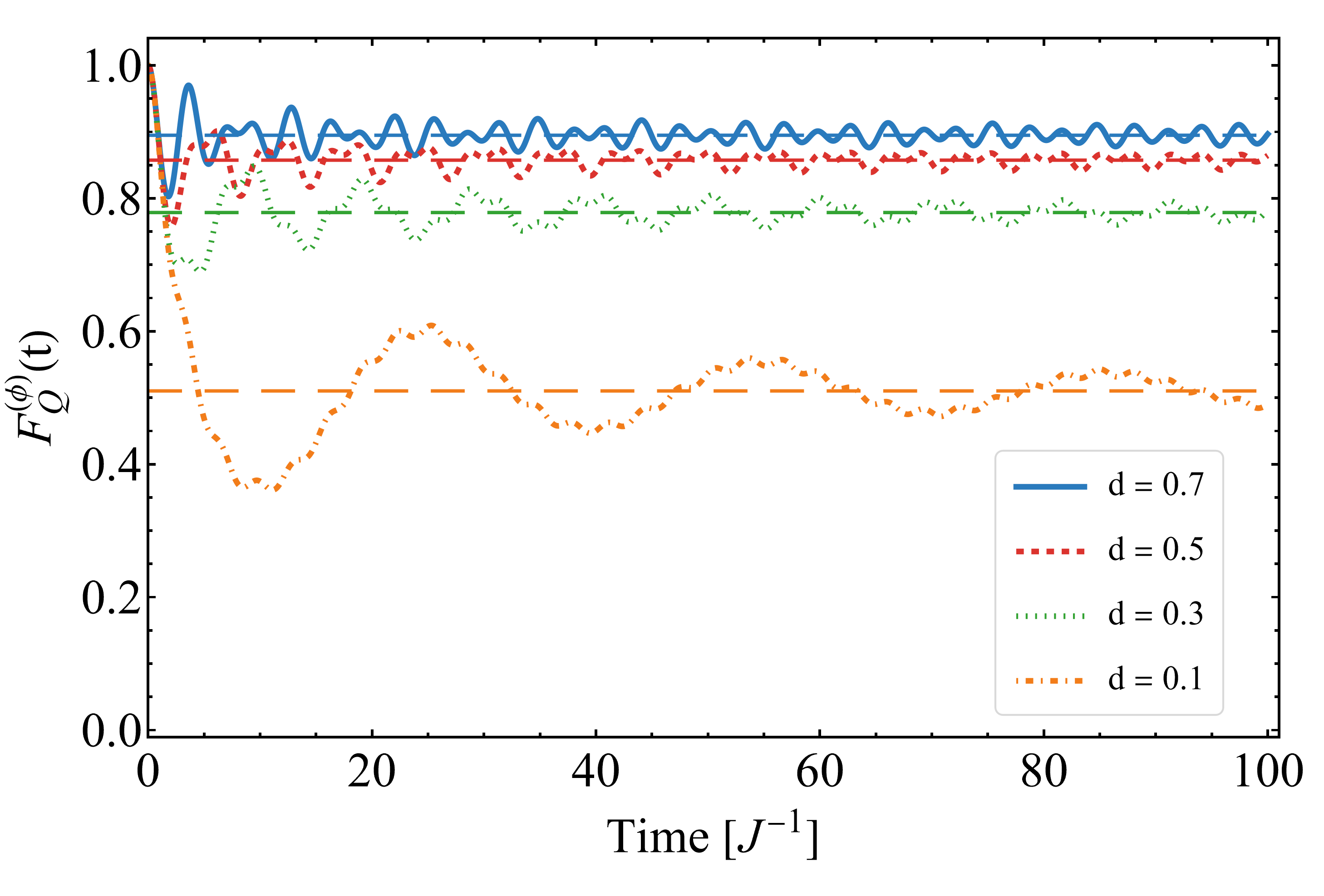}
    \caption{Time-dependent phase-QFI $F_Q^{(\phi)}(t)$ at $\Delta=0$ and $g=0.4J$ for four representative values of the dimerization parameter, $d=0.7$, $0.5$, $0.3$, and $0.1$. The horizontal lines indicate the analytic central in-gap bound-state prediction derived below [see Eq.~\eqref{eq:qfi_steady_state_analytic}]. Numerical results are obtained for a finite SSH chain with $2L+1$ unit cells, $L=500$, using a Krylov subspace of dimension $350$.}
    \label{fig:qfi_vs_dimerization}
\end{figure}

We begin with the dependence on the SSH dimerization at zero detuning. This is the cleanest setting in which to isolate the role of the central gap, because the bare emitter frequency is pinned to the gap center and the long-time phase sensitivity is governed directly by how strongly the dimerized lattice suppresses relaxation. Figure~\ref{fig:qfi_vs_dimerization} shows the time evolution of $F_Q^{(\phi)}(t)$ for $\Delta=0$, fixed coupling strength $g=0.4J$, and four representative values of the dimerization parameter, $d=0.7$, $0.5$, $0.3$, and $0.1$. The horizontal lines indicate the corresponding central in-gap bound-state predictions discussed below.

The trend in Fig.~\ref{fig:qfi_vs_dimerization} is clear: stronger dimerization produces a larger retained phase-QFI at late times. Physically, increasing $|d|$ widens the central gap and strengthens the gap-supported incomplete relaxation of the emitter. As $|d|$ is reduced, the gap narrows, the nondecaying component of the emitter dynamics becomes smaller, and the long-time metrological signal is progressively suppressed. The dimerization parameter therefore acts as a direct microscopic control knob for long-time phase-information retention.

The same parameter also affects the transient dynamics. For all values of $d$ shown, the initial decay is followed by damped oscillations before the curves settle into their late-time behavior. These oscillations are the metrological signature of the underlying non-Markovian emitter dynamics induced by the structured SSH spectrum. Their visibility changes with $d$ because the same spectral reorganization that controls the retained late-time signal also modifies the interference between bound-state and continuum contributions during the transient regime.

At resonance, the retained late-time behavior can be characterized analytically. The phase-QFI associated with the nondecaying component is determined by the corresponding emitter amplitude,
\begin{equation}
F_Q^{(\phi)}(\infty)=|u(\infty)|^2,
\label{eq:qfi_steady_state}
\end{equation}
and for $\Delta=0$ one obtains from the bound-state contribution \cite{Bello2019}
\begin{equation}
F_Q^{(\phi)}(\infty)=
\left(1+\frac{g^2}{4J^2 |d|}\right)^{-2}.
\label{eq:qfi_steady_state_analytic}
\end{equation}
Equation~\eqref{eq:qfi_steady_state_analytic} shows directly how the retained resonant signal is controlled by the SSH gap scale. Any nonzero dimerization opens a finite central gap and therefore yields a finite retained phase-QFI, whereas the limit $|d|\to 0$ closes the gap and drives the late-time metrological signal to zero. The horizontal lines in Fig.~\ref{fig:qfi_vs_dimerization} correspond to this expression.

\begin{figure}[tbp]
    \centering
    \includegraphics[width=\columnwidth]{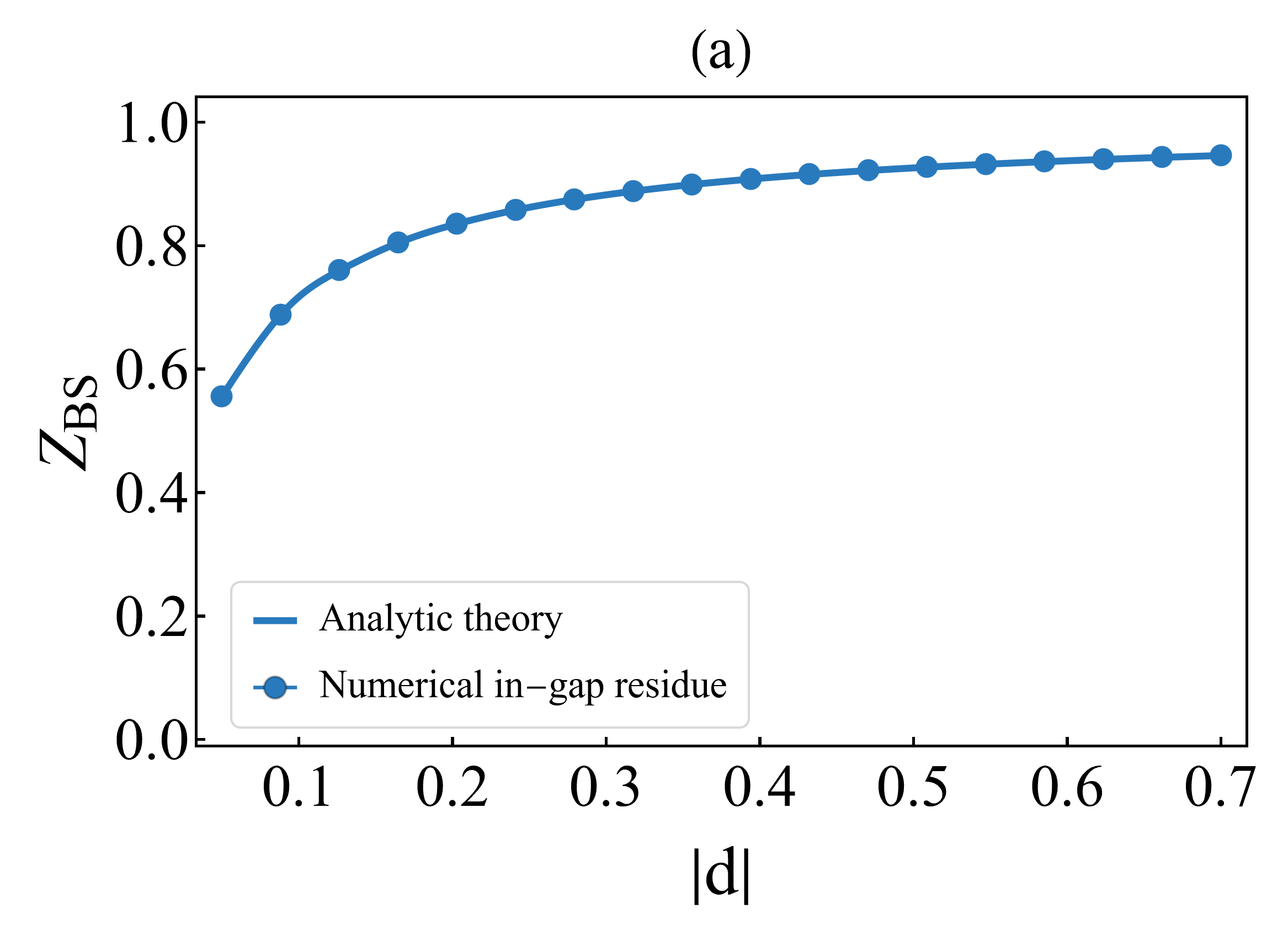}\\
    \includegraphics[width=\columnwidth]{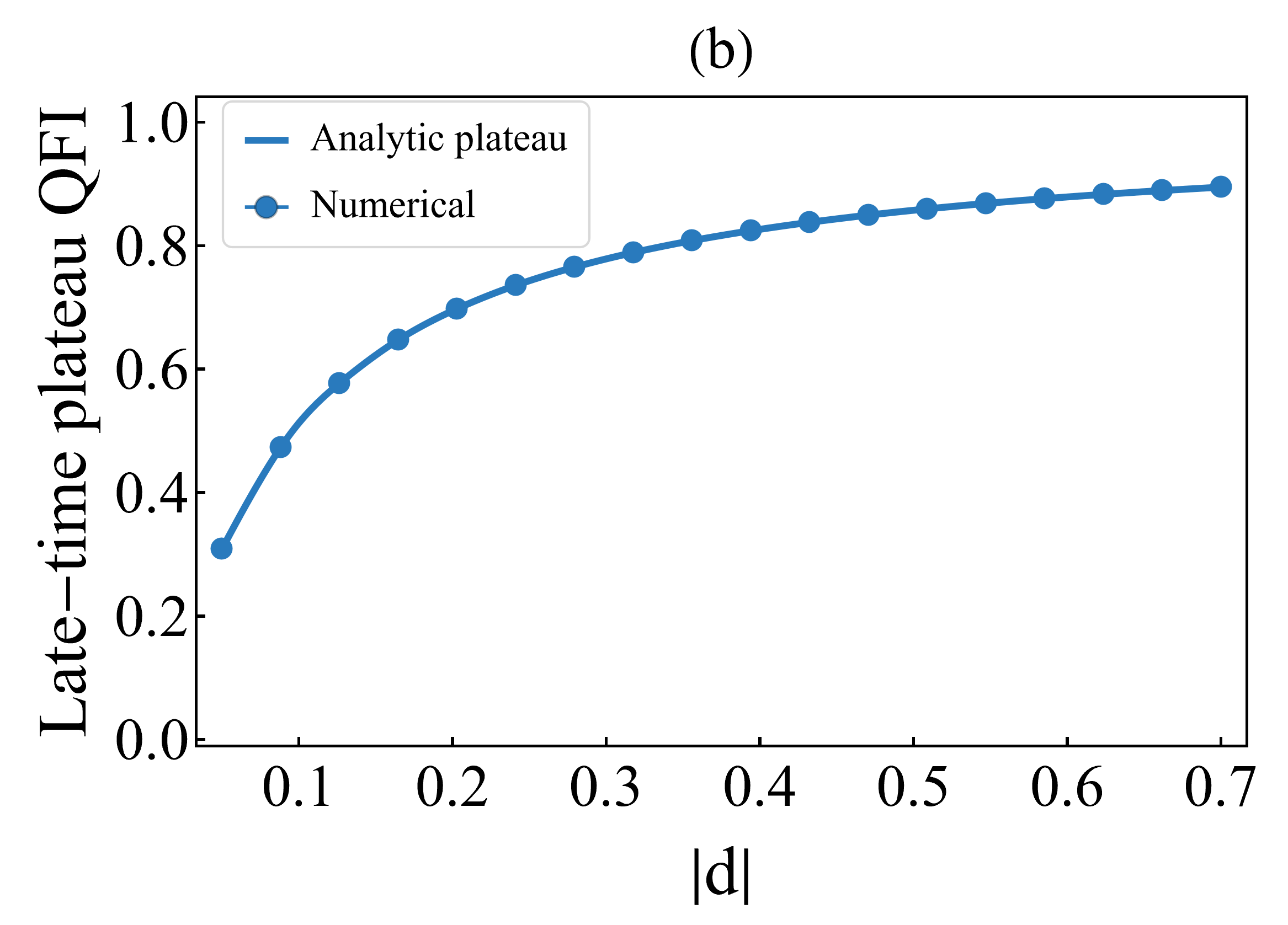}
    \caption{Comparison between numerical and analytic bound-state diagnostics at $\Delta=0$ and $g=0.4J$. (a) Emitter bound-state residue $Z_{\mathrm{BS}}$ extracted from the in-gap eigenstate of a finite SSH chain, together with the analytic expression in Eq.~\eqref{eq:zbs_analytic}. (b) Retained phase-QFI obtained from the central in-gap bound-state relation $F_Q^{(\phi)}=Z_{\mathrm{BS}}^2$, together with the analytic result in Eq.~\eqref{eq:qfi_steady_state_analytic}. Numerical diagonalizations use a finite chain with $2L+1$ unit cells and $L=220$.}
    \label{fig:bound_state_residue}
\end{figure}

The microscopic origin of the resonant signal can be made explicit through the bound-state content of the coupled emitter-bath Hamiltonian. For $\Delta=0$ and $d\neq 0$, the finite SSH chain contains a single in-gap eigenstate in the present bulk local-coupling geometry, with emitter overlap
\begin{equation}
Z_{\mathrm{BS}}=\sum_{\alpha\in \mathrm{gap}} |\langle e|\psi_\alpha\rangle|^2,
\label{eq:zbs_def}
\end{equation}
controlling the nondecaying part of the survival amplitude. Figure~\ref{fig:bound_state_residue} shows that the numerical in-gap residue coincides with the analytic prediction
\begin{equation}
Z_{\mathrm{BS}}=\left(1+\frac{g^2}{4J^2 |d|}\right)^{-1},
\label{eq:zbs_analytic}
\end{equation}
so that
\begin{equation}
F_Q^{(\phi)}(\infty)=Z_{\mathrm{BS}}^2.
\label{eq:qfi_zbs_relation}
\end{equation}
The retained resonant signal is therefore fixed directly by the emitter weight carried by the in-gap bound state.

\begin{figure}[tbp]
    \centering
    \includegraphics[width=\columnwidth]{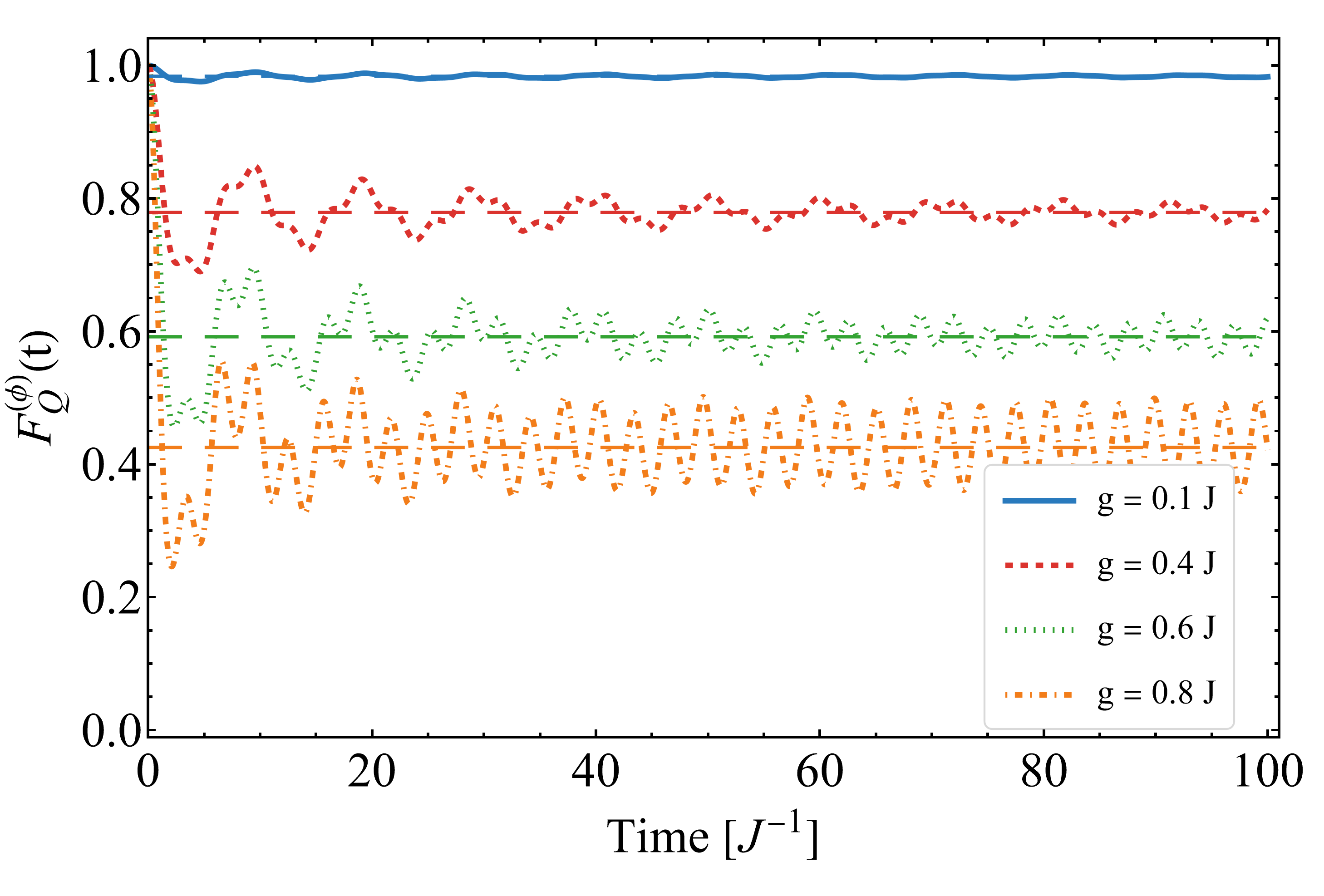}
    \caption{Time-dependent phase-QFI $F_Q^{(\phi)}(t)$ at zero detuning, $\Delta=0$, and fixed dimerization $d=0.3$ for four representative emitter-bath coupling strengths, $g/J=0.1$, $0.4$, $0.6$, and $0.8$. The horizontal lines indicate the corresponding central in-gap bound-state predictions from Eq.~\eqref{eq:qfi_steady_state_analytic}. Numerical results are obtained for a finite SSH chain with $2L+1$ unit cells, $L=500$, using a Krylov subspace of dimension $350$.}
    \label{fig:qfi_vs_coupling}
\end{figure}

It is also useful to examine the role of the emitter-bath coupling strength at resonance. Figure~\ref{fig:qfi_vs_coupling} shows the time evolution of $F_Q^{(\phi)}(t)$ at fixed dimerization $d=0.3$ and zero detuning for several representative values of $g$. In the present bulk SSH geometry, increasing the coupling strength does not enhance the retained late-time metrological signal. Instead, Eqs.~\eqref{eq:zbs_analytic} and \eqref{eq:qfi_zbs_relation} show that, at fixed $d$, both the emitter bound-state residue and the retained phase-QFI decrease monotonically with $g$. Physically, stronger hybridization with the bath transfers more emitter weight away from the nondecaying in-gap component, even though the gap remains open and the bound state persists. At the same time, the transient dynamics is reorganized: the initial decay and the oscillatory structure become more pronounced as $g$ increases, indicating stronger interference between bound-state and continuum contributions before the late-time regime is reached. The coupling strength therefore controls not only the retained phase information but also the character of the transient approach to the late-time regime.

\begin{figure}[tbp]
    \centering
    \includegraphics[width=\columnwidth]{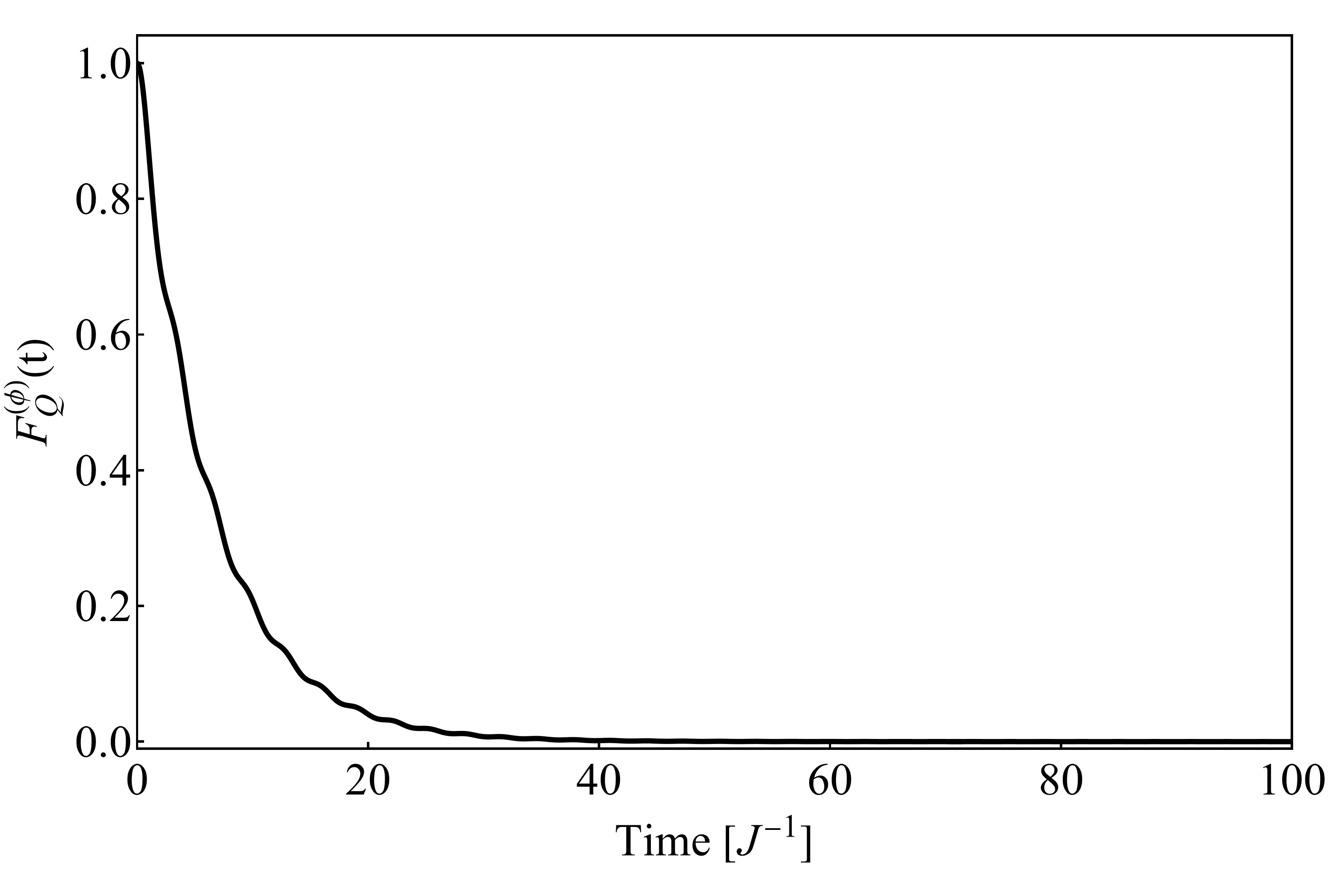}
    \caption{Time-dependent phase-QFI $F_Q^{(\phi)}(t)$ in the uniform-chain limit $d=0$ at $\Delta=0$ and $g=0.4J$. Numerical results are obtained for a finite SSH chain with $2L+1$ unit cells, $L=500$, using a Krylov-space propagation of dimension $350$.}
    \label{fig:qfi_uniform_control}
\end{figure}

This interpretation is reinforced by the uniform-chain control shown in Fig.~\ref{fig:qfi_uniform_control}. In the gapless limit $d=0$, the SSH lattice reduces to a uniform nearest-neighbor chain and the central gap closes. Relative to the gapped cases in Fig.~\ref{fig:qfi_vs_dimerization}, the retained late-time signal disappears and the phase-QFI decays essentially to zero over the same observation window. This confirms, within the same Hamiltonian family, that the persistent late-time signal for $d\neq 0$ is not a generic feature of resonant emitter dynamics but a direct consequence of gap-supported incomplete relaxation in the dimerized SSH lattice.

A final point concerns the sign of the SSH dimerization. In the present bulk local-coupling geometry, the local phase-QFI does not distinguish between $d$ and $-d$: the retained late-time phase-QFI is an even function of $d$, and numerically the full dynamics is invariant under $d\to -d$ within plotting resolution. Thus the single-emitter quantity $F_Q^{(\phi)}(t)$ probes the presence and size of the spectral gap, together with the associated bound-state support, but not the winding sector itself. This is consistent with the underlying bulk SSH dynamics, where the sign of $d$ affects the spatial chirality of the photonic cloud but not the local survival probability at the emitter site \cite{Bello2019}. The present protocol therefore resolves gap physics and bound-state support, while leaving explicitly topology-sensitive metrological signatures to more elaborate geometries \cite{Vega2023,TopoSensor2025}.

\subsection{Central in-gap bound state at arbitrary detuning}
\label{sec:analytic_offresonant}

The resonant analysis above showed that the retained phase-QFI is controlled by the emitter weight carried by the in-gap bound state, and that this explains both the dimerization dependence of the late-time signal and its suppression as the coupling strength $g$ is increased at fixed $d$. We now show that this bound-state interpretation extends away from resonance. In the present bulk local-coupling geometry, the emitter-bath system supports exactly one bound state inside the central SSH gap for every real detuning. Its residue provides the central analytical quantity used to organize the off-resonant dynamics below.

In the single-excitation sector, the retarded Green's function of the emitter obeys the Dyson equation
\begin{equation}
G(\omega)=\frac{1}{\omega-\Delta-g^2 G_A^0(\omega)},
\label{eq:dyson}
\end{equation}
where $G_A^0(\omega)$ is the local $A$-site retarded Green's function of the bare SSH bath, defined as the $\eta\to 0^+$ limit of $G_A^0(\omega+i\eta)$. Because the SSH Bloch eigenstates satisfy $|\langle A|\psi_\pm(k)\rangle|^2=1/2$ for all $k$ and for both bands, one has
\begin{equation}
G_A^0(\omega)
=\lim_{\eta\to 0^+}\int_{-\pi}^{\pi}\frac{dk}{2\pi}\,\frac{\omega+i\eta}{(\omega+i\eta)^2-\omega_+^2(k)},
\label{eq:GA0_integral}
\end{equation}
with $\omega_+(k)$ given by Eq.~\eqref{eq:ssh_dispersion}. For real frequencies inside the central gap, $|\omega|<2J|d|$, the bath spectral density vanishes and the integral is purely real. Evaluating it in closed form yields
\begin{equation}
G_A^0(\omega)=-\frac{\omega}{\sqrt{(4J^2-\omega^2)(4J^2 d^2-\omega^2)}},
\label{eq:GA0_closed}
\end{equation}
and
\begin{equation}
\frac{dG_A^0(\omega)}{d\omega}
=-\frac{16J^4 d^2-\omega^4}{\left[(4J^2-\omega^2)(4J^2 d^2-\omega^2)\right]^{3/2}}.
\label{eq:GA0_derivative}
\end{equation}

Throughout the central gap, $G_A^0(\omega)$ is real, analytic, and strictly decreasing, with $G_A^0(\omega)\to\mp\infty$ as $\omega\to\pm 2J|d|$ from inside. It follows that the in-gap pole condition
\begin{equation}
\omega_{\mathrm{BS}}-\Delta-g^2 G_A^0(\omega_{\mathrm{BS}})=0
\label{eq:pole_equation}
\end{equation}
has exactly one root $\omega_{\mathrm{BS}}(\Delta)$ in the central gap for every $\Delta\in\mathbb{R}$. The corresponding emitter residue is
\begin{equation}
Z_{\mathrm{BS}}(\Delta)
=\left[1-g^2\,\frac{dG_A^0}{d\omega}\bigg|_{\omega_{\mathrm{BS}}}\right]^{-1},
\label{eq:ZBS_residue_general}
\end{equation}
which satisfies $Z_{\mathrm{BS}}(\Delta)\in(0,1)$ throughout the gap. At exact resonance, $\omega_{\mathrm{BS}}(0)=0$, and Eq.~\eqref{eq:ZBS_residue_general} reduces to the resonant expression in Eq.~\eqref{eq:zbs_analytic}. In particular, the monotonic decrease of the resonant residue with increasing $g$ at fixed $d$, discussed in Fig.~\ref{fig:qfi_vs_coupling}, follows immediately from this closed-form result.

The pole at $\omega_{\mathrm{BS}}$ defines the central in-gap bound-state contribution to the emitter survival amplitude,
\begin{equation}
u_{\mathrm{BS}}(t)=Z_{\mathrm{BS}}(\Delta)\,e^{-i\omega_{\mathrm{BS}}(\Delta)t}.
\label{eq:u_decomposition}
\end{equation}
The remaining part of the survival amplitude contains the contributions from the two SSH bands, which dephase in the thermodynamic limit, and, in a finite-band problem, may also contain additional discrete contributions outside the outer band edges. We checked the finite-chain spectra used in the numerical calculations below and found that, for the parameter regimes shown, such additional states, when present, carry negligible emitter weight on the scale of the reported late-time indicators. The post-transient signal is therefore controlled by the central in-gap pole. The corresponding bound-state contribution to the retained phase-QFI is
\begin{equation}
F_{Q,\mathrm{BS}}^{(\phi)}=Z_{\mathrm{BS}}(\Delta)^2,
\label{eq:FQ_ZBS_exact}
\end{equation}
which is used below as the central bound-state benchmark for the off-resonant dynamics.

This result has three immediate consequences for the present setup. First, it extends the resonant bound-state interpretation of the retained phase-QFI to arbitrary detuning. Second, since both the in-gap pole equation and the residue depend on $d$ only through $d^2$, the sign independence of the local phase-QFI follows from the bulk local-coupling geometry rather than from a numerical coincidence. Third, as the bound-state energy approaches a band edge, the derivative in Eq.~\eqref{eq:GA0_derivative} diverges and the residue vanishes, $Z_{\mathrm{BS}}\to 0$, anticipating the suppression of the late-time metrological signal near the gap edge.

With this bound-state benchmark in hand, we now turn to the off-resonant numerics and use it to interpret the operational indicators introduced below.

\subsection{Off-resonant crossover and operational diagnostics}
\label{sec:results_offresonant}

\begin{figure}[!htbp]
    \centering
    \includegraphics[width=\columnwidth]{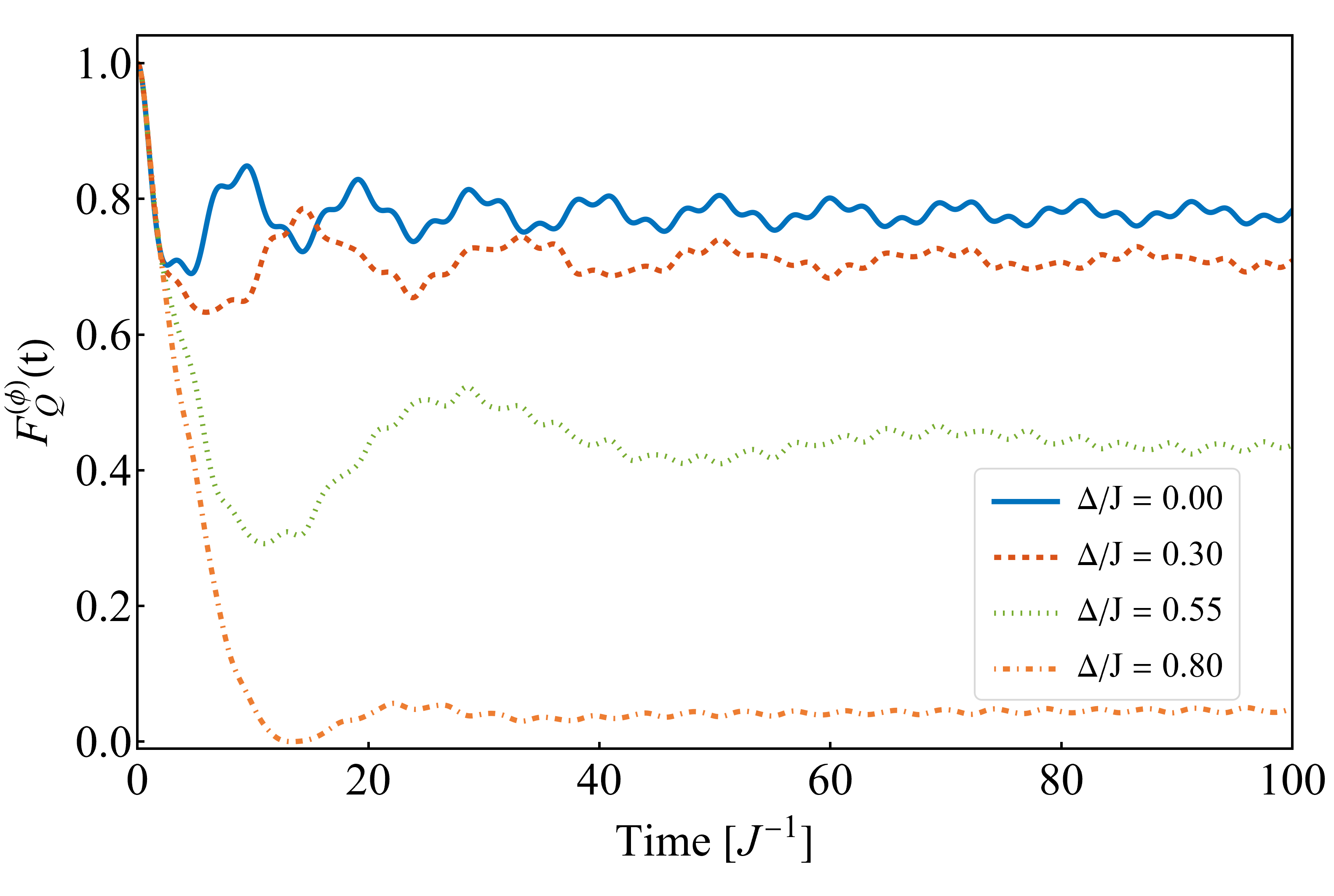}
    \caption{Time-dependent phase-QFI $F_Q^{(\phi)}(t)$ at fixed dimerization $d=0.3$ and coupling strength $g=0.4J$ for four representative detunings, $\Delta/J=0$, $0.3$, $0.55$, and $0.8$. Numerical results are obtained for a finite SSH chain with $2L+1$ unit cells, $L=500$, and a Krylov subspace of dimension $350$.}
    \label{fig:qfi_vs_detuning}
\end{figure}

We now turn to the off-resonant regime, where detuning moves the emitter across the SSH spectrum at fixed lattice dimerization. Whereas the resonant analysis above isolated how the gap width and the coupling strength control the retained late-time signal, the detuning scan shows how metrological retention is reorganized as the bare emitter frequency is displaced from the gap center toward the band edge and then into the continuum. Figure~\ref{fig:qfi_vs_detuning} shows $F_Q^{(\phi)}(t)$ for $d=0.3$, $g=0.4J$, and four representative detunings, $\Delta/J=0$, $0.3$, $0.55$, and $0.8$, corresponding respectively to the gap center, an off-center in-gap point, a near-edge regime, and a continuum frequency.

The crossover is physically sharp. At the gap center, the phase-QFI remains strongly protected and approaches a sizable late-time value. Moving away from resonance while remaining inside the gap weakens this protection but still leaves a substantial residual signal. Near the band edge, the dynamics becomes much more sensitive to spectral details, and the QFI reflects a more delicate competition between transient oscillations and late-time decay. Once the bare emitter frequency enters the continuum, the late-time metrological protection is strongly reduced. The detuning dependence is therefore not a small perturbation around the resonant case. It provides a direct microscopic interpolation between gap-protected retention and continuum-dominated relaxation within the same SSH reservoir.

For the bulk single-emitter geometry considered here, the phase-QFI is numerically even in the detuning to machine precision, so that $F_Q^{(\phi)}(t;\Delta)=F_Q^{(\phi)}(t;-\Delta)$ in all cases examined. It is therefore sufficient to display only nonnegative detunings in Fig.~\ref{fig:qfi_vs_detuning}. Together with the $d^2$ dependence of the in-gap pole equation and residue discussed above, this shows that the local metrological response is controlled by the spectral position of the emitter relative to the gap and the band edges, rather than by the sign of either control parameter in the present symmetric bulk geometry.

Although Eq.~\eqref{eq:FQ_ZBS_exact} provides the central bound-state benchmark, the numerically accessible off-resonant dynamics does not always settle into an easily identifiable plateau on finite time windows. For this reason, it is useful to introduce operational late-time indicators that summarize the post-transient behavior without forcing a plateau interpretation where residual oscillations remain visible. The first such quantity is the late-time averaged phase-QFI,
\begin{equation}
\overline{F}_Q^{(\phi)}=\frac{1}{t_2-t_1}\int_{t_1}^{t_2}F_Q^{(\phi)}(t)\,dt,
\label{eq:late_time_average_qfi}
\end{equation}
evaluated over a finite observation window $[t_1,t_2]$. In the calculations below we take $t_1=40J^{-1}$ and $t_2=100J^{-1}$. This window probes the post-transient regime while remaining well within the pre-recurrence dynamics of the finite chains used here. Numerically, the same quantity can also be evaluated directly from the Krylov spectral data through the finite-window kernel associated with the reduced tridiagonal spectrum, providing a useful consistency check against direct time sampling. The quantity $\overline{F}_Q^{(\phi)}$ should therefore be viewed as an operational late-time indicator rather than as a substitute for an exact infinite-time limit.

In practice, $\overline{F}_Q^{(\phi)}$ captures the same crossover already visible in the time-domain analysis: comparatively large values inside the gap, a marked reduction near the band edge, and strongly suppressed late-time response in the continuum. It therefore provides a natural bridge between the full time traces and the compact parameter-space maps discussed below.

\begin{figure}[!htbp]
    \centering
    \includegraphics[width=\columnwidth]{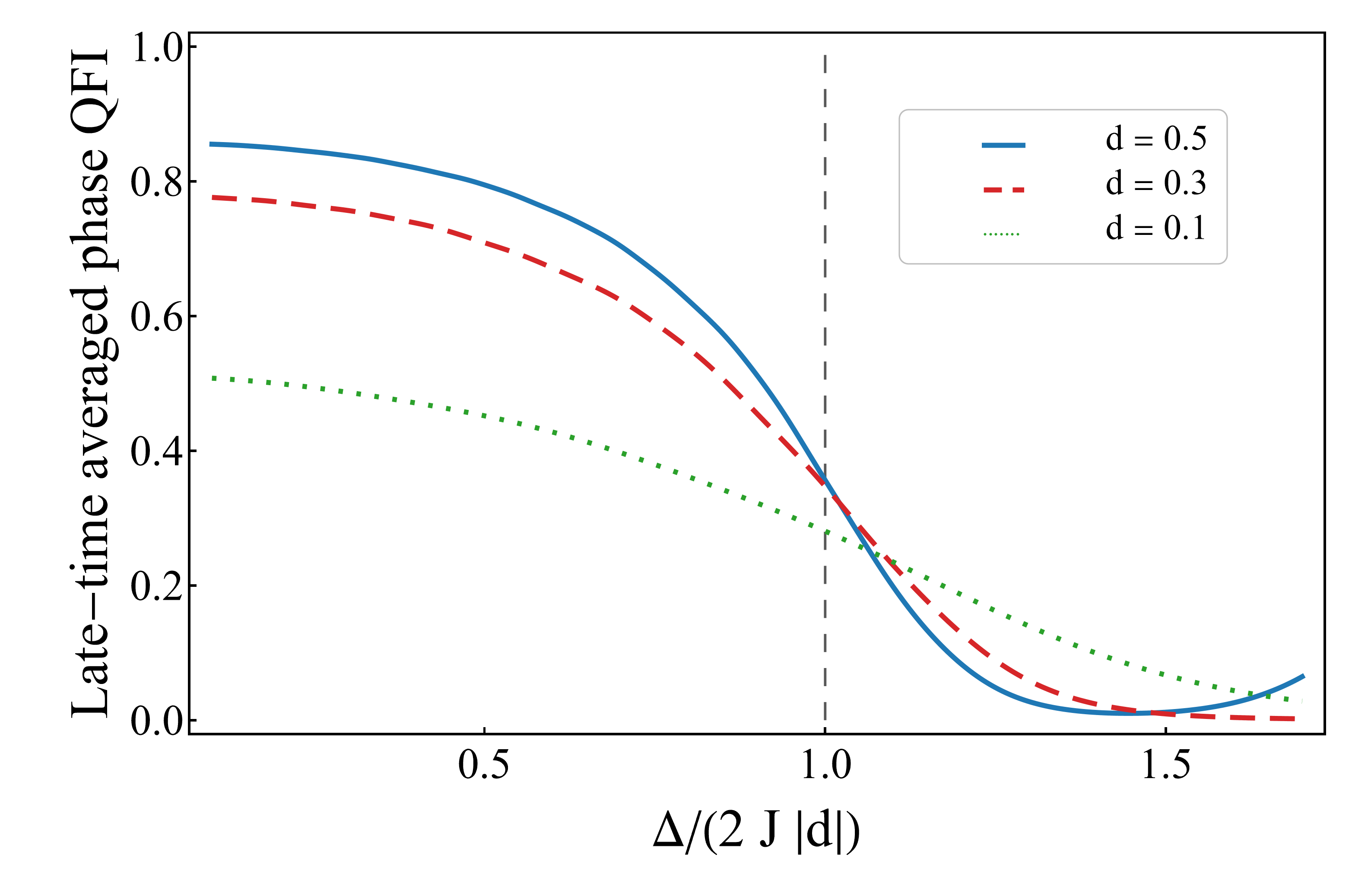}
    \caption{Late-time averaged phase-QFI $\overline{F}_Q^{(\phi)}$ as a function of the normalized detuning $\Delta/(2J|d|)$ for three representative dimerizations, $d=0.5$, $0.3$, and $0.1$, at fixed coupling strength $g=0.4J$. The average is evaluated over the time window $40J^{-1}\le t \le 100J^{-1}$. The vertical dashed line marks the band-edge condition $\Delta/(2J|d|)=1$. Numerical results are obtained for a finite SSH chain with $2L+1$ unit cells, $L=220$, via exact diagonalization of the single-excitation Hamiltonian.}
    \label{fig:qfi_gap_normalized}
\end{figure}

A useful way to organize this late-time behavior is to measure the detuning relative to the SSH gap edge through the normalized variable $\Delta/(2J|d|)$. Figure~\ref{fig:qfi_gap_normalized} shows the late-time averaged phase-QFI as a function of this quantity for three representative dimerizations, $d=0.5$, $0.3$, and $0.1$, at fixed coupling strength $g=0.4J$. This representation addresses a sharper physical question: to what extent is the late-time metrological response governed by the emitter position relative to the gap edge, rather than by the bare values of $d$ and $\Delta$ separately?

Presented in this way, the late-time metrological response is seen to be governed to a large extent by the emitter position relative to the gap edge. Deep inside the gap, $\Delta/(2J|d|)<1$, the late-time averaged phase-QFI remains comparatively large for all values of $d$ shown here. As the normalized detuning approaches unity, the averaged QFI drops, and once the bare emitter frequency moves into the continuum, $\Delta/(2J|d|)>1$, the late-time metrological protection is progressively reduced. The normalized-detuning representation therefore separates, in a compact way, the gap-dominated, edge-sensitive, and continuum-dominated regimes.

The curves do not collapse exactly onto a single universal function of $\Delta/(2J|d|)$. This indicates that the gap width is not the only quantity controlling the late-time metrological behavior. Additional spectral details, including the precise structure of the emitter-environment coupling and the redistribution of spectral weight between bound-state and continuum sectors, remain relevant. Even so, the normalized detuning provides a useful organizing variable: much of the crossover in the late-time phase-QFI is set by the ratio of the emitter frequency to the SSH gap edge, rather than by the bare detuning alone. This noncollapse is conceptually important, because it shows that the SSH bath is not reduced here to a one-parameter effective-gap model. Rather, the late-time metrological response reflects the full microscopic spectral structure seen by the emitter.

A complementary operational summary is provided by the QFI retention time $t_{\eta}$, defined as the first time at which the phase-QFI falls below a prescribed threshold value $\eta$,
\begin{equation}
F_Q^{(\phi)}(t_\eta)=\eta.
\label{eq:qfi_retention_time}
\end{equation}
In the calculations below we choose $\eta=0.2$. This is not a universal metrological benchmark. Rather, it provides a simple diagnostic for comparing different detunings and gap sizes within the same protocol. Moderate changes of the threshold shift the numerical values of $t_\eta$ but do not alter the qualitative ordering across the gap, band-edge, and continuum regimes.

\begin{figure}[!htbp]
    \centering
    \includegraphics[width=\columnwidth]{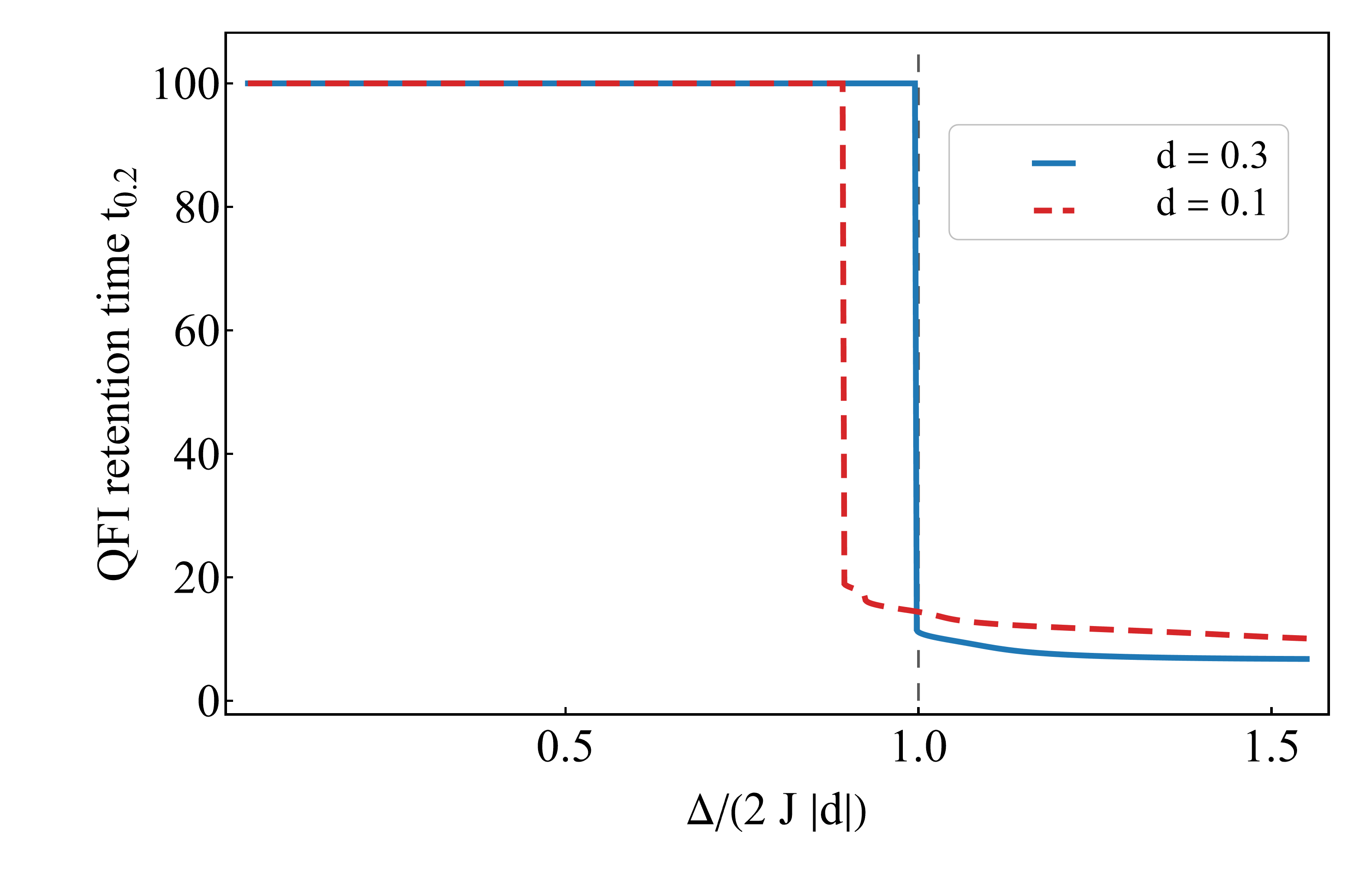}
    \caption{QFI retention time $t_{0.2}$ as a function of the normalized detuning $\Delta/(2J|d|)$ for two representative dimerizations, $d=0.3$ and $d=0.1$, at fixed coupling strength $g=0.4J$. The vertical dashed line marks the band-edge condition $\Delta/(2J|d|)=1$. Numerical results are obtained for a finite SSH chain with $2L+1$ unit cells, $L=520$, and a Krylov subspace of dimension $360$.}
    \label{fig:qfi_retention_time}
\end{figure}

Figure~\ref{fig:qfi_retention_time} provides an operational readout of the same crossover identified in the time-domain and late-time-averaged analyses. Deep inside the gap, the phase-QFI stays above the chosen threshold for long intervals and the corresponding retention time is large. As the normalized detuning approaches the band edge, the retention time decreases, while in the continuum it drops rapidly. The comparison between $d=0.3$ and $d=0.1$ further shows that a narrower gap leads to a more fragile metrological response even when the detuning is measured relative to the gap edge.

Unlike the trivial maximization of the QFI over all times, which is always attained at $t=0$ for the present phase-encoded initial state, the retention time probes a genuinely dynamical aspect of metrological performance. It therefore complements the late-time average and reinforces the conclusion that the SSH spectral structure controls not only the magnitude but also the temporal persistence of metrologically useful information.

\begin{figure}[!htbp]
    \centering
    \includegraphics[width=0.96\columnwidth]{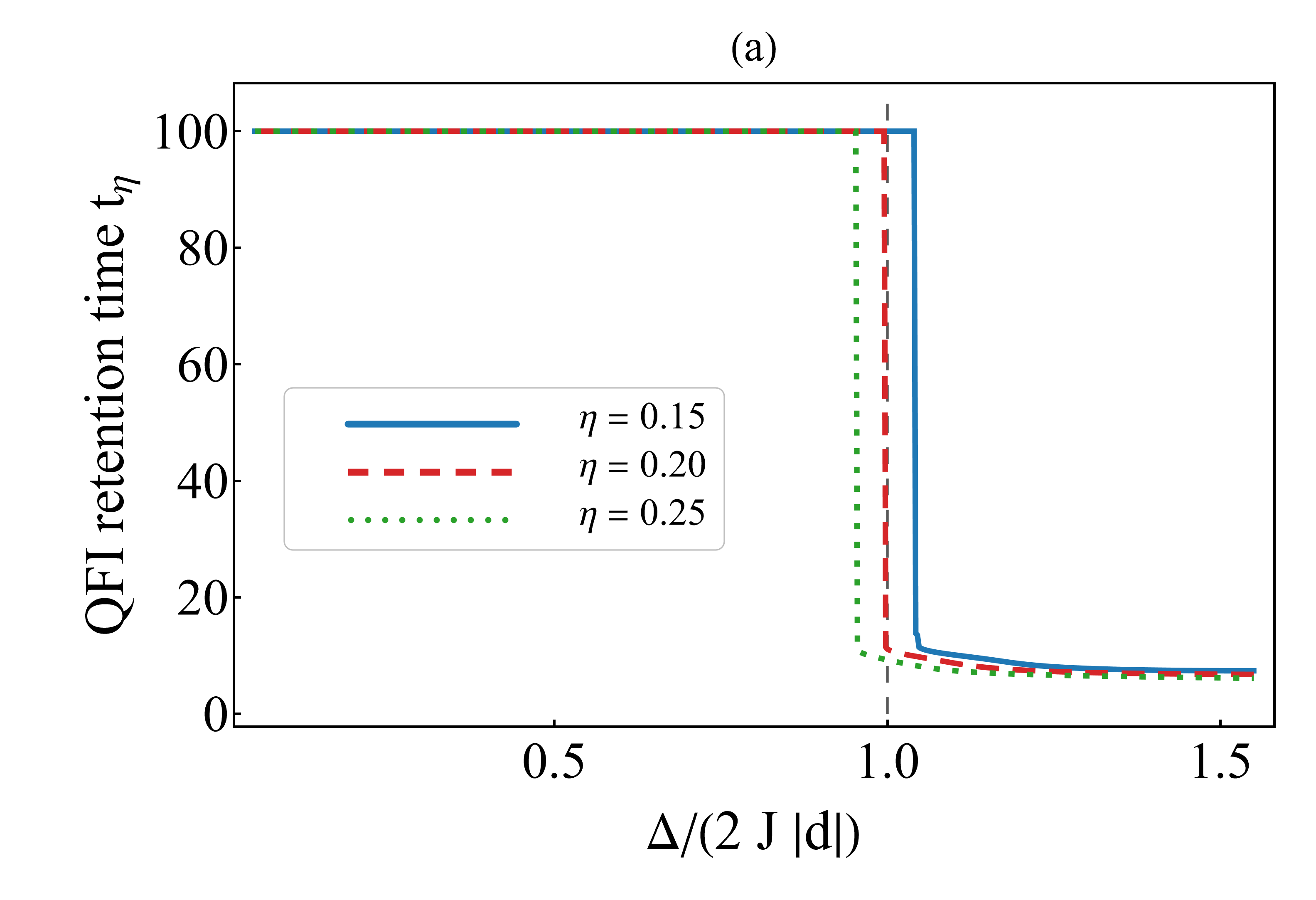}\\
    \includegraphics[width=0.96\columnwidth]{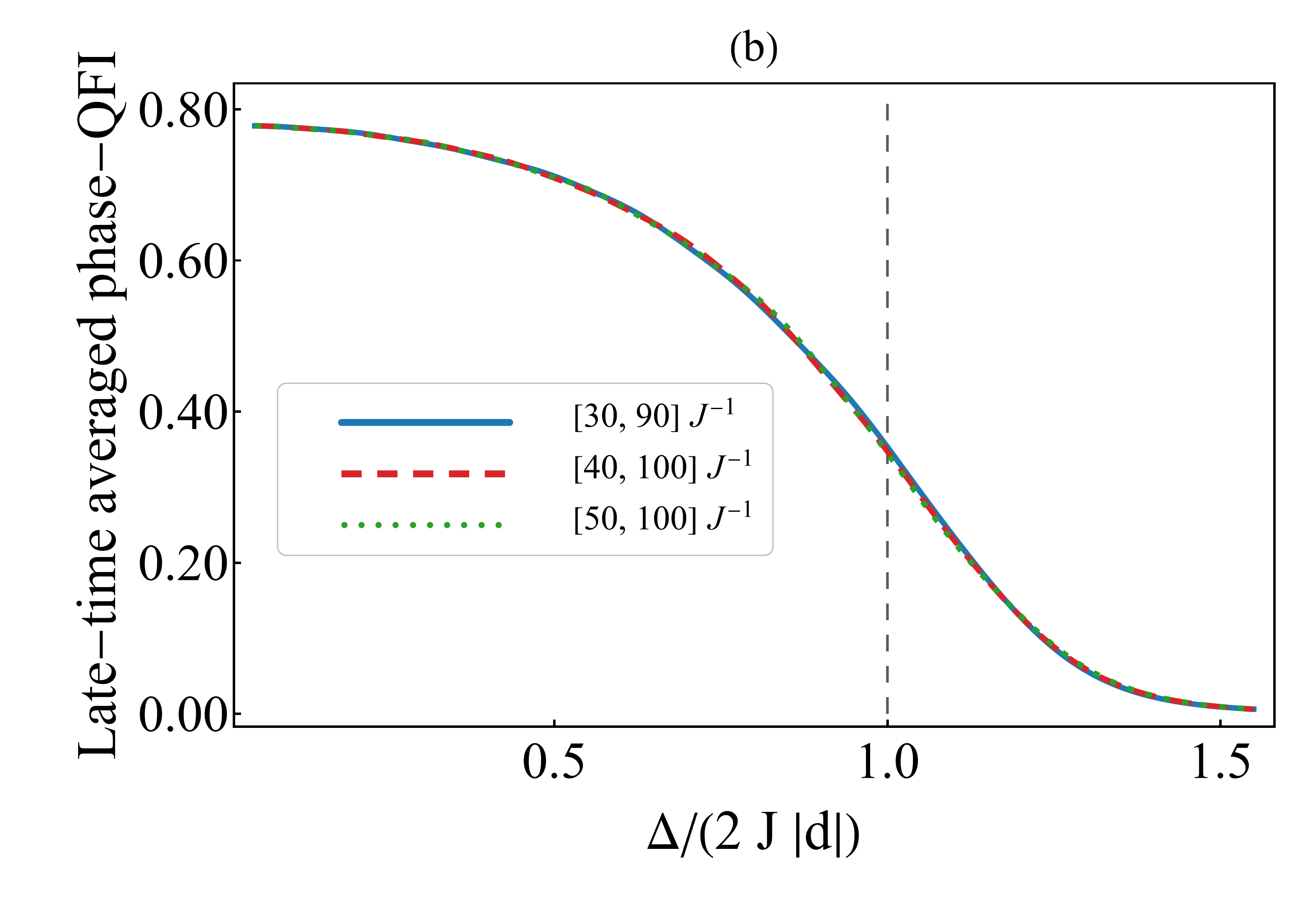}\\
    \caption{Robustness checks for the operational indicators used in the off-resonant analysis at fixed $d=0.3$ and $g=0.4J$. (a) QFI retention time $t_\eta$ versus normalized detuning for three threshold values, $\eta=0.15$, $0.2$, and $0.25$. Whenever the threshold is not crossed before the end of the simulation window, the retention time is capped at $t_{\max}=100J^{-1}$. (b) Late-time averaged phase-QFI $\overline{F}_Q^{(\phi)}$ versus normalized detuning for three averaging windows, $[30,90]J^{-1}$, $[40,100]J^{-1}$, and $[50,100]J^{-1}$. The vertical dashed line marks the band-edge condition $\Delta/(2J|d|)=1$. Numerical results are obtained for a finite SSH chain with $2L+1$ unit cells for $L=520$ with a Krylov subspace of dimension $360$.}
    \label{fig:operational_robustness}
\end{figure}

Because both $\overline{F}_Q^{(\phi)}$ and $t_\eta$ involve finite operational choices, it is important to verify explicitly that the physical conclusions do not depend sensitively on a single numerical convention. Figure~\ref{fig:operational_robustness} summarizes two representative robustness checks for the off-resonant analysis at fixed $d=0.3$ and $g=0.4J$. Panel~(a) compares the retention-time trends obtained for three nearby thresholds, $\eta=0.15$, $0.2$, and $0.25$. Panel~(b) compares the gap-normalized late-time averaged QFI obtained from three different averaging windows, $[30,90]J^{-1}$, $[40,100]J^{-1}$, and $[50,100]J^{-1}$.

These checks support the main interpretation of the off-resonant results. The operational indicators are not intended as context-free universal definitions of metrological performance. They are compact diagnostics of how long useful phase information survives in the present dissipative protocol. The fact that the same parameter ordering persists under moderate variations of both the threshold and the averaging window shows that the observed trends are controlled by the SSH spectral crossover itself, not by a fine-tuned numerical convention.

The late-time average and the retention time already provide compact operational summaries. A third, complementary indicator isolates the portion of the evolution that remains metrologically useful after the initial transient. Because the protocol always starts from $F_Q^{(\phi)}(0)=1$, maximizing the QFI over all times does not distinguish between parameter sets. A more informative quantity is the post-transient useful-window duration
\begin{equation}
W_{\eta}(t_{\mathrm{cut}},T)=\int_{t_{\mathrm{cut}}}^{T} dt\; \Theta\!\left[F_Q^{(\phi)}(t)-\eta\right],
\label{eq:useful_window}
\end{equation}
where $\Theta$ is the Heaviside step function, $t_{\mathrm{cut}}$ excludes the preparation transient, and $\eta$ is an operational phase-sensitivity threshold. In contrast to $t_\eta$, Eq.~\eqref{eq:useful_window} measures the total amount of late-time experimental time during which the signal remains above a chosen target.

\begin{figure}[tbp]
    \centering
    \includegraphics[width=\columnwidth]{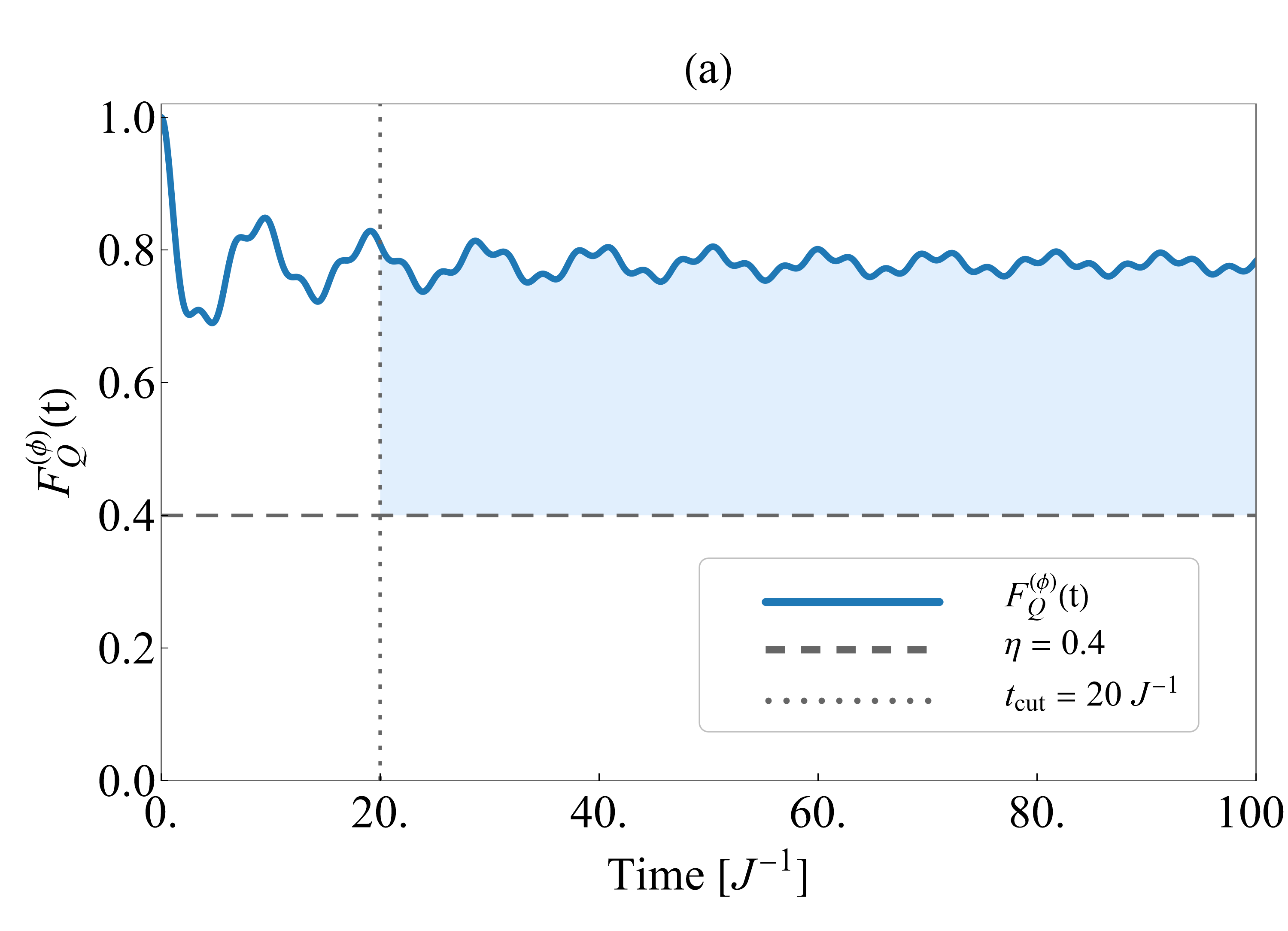}\\
    \includegraphics[width=\columnwidth]{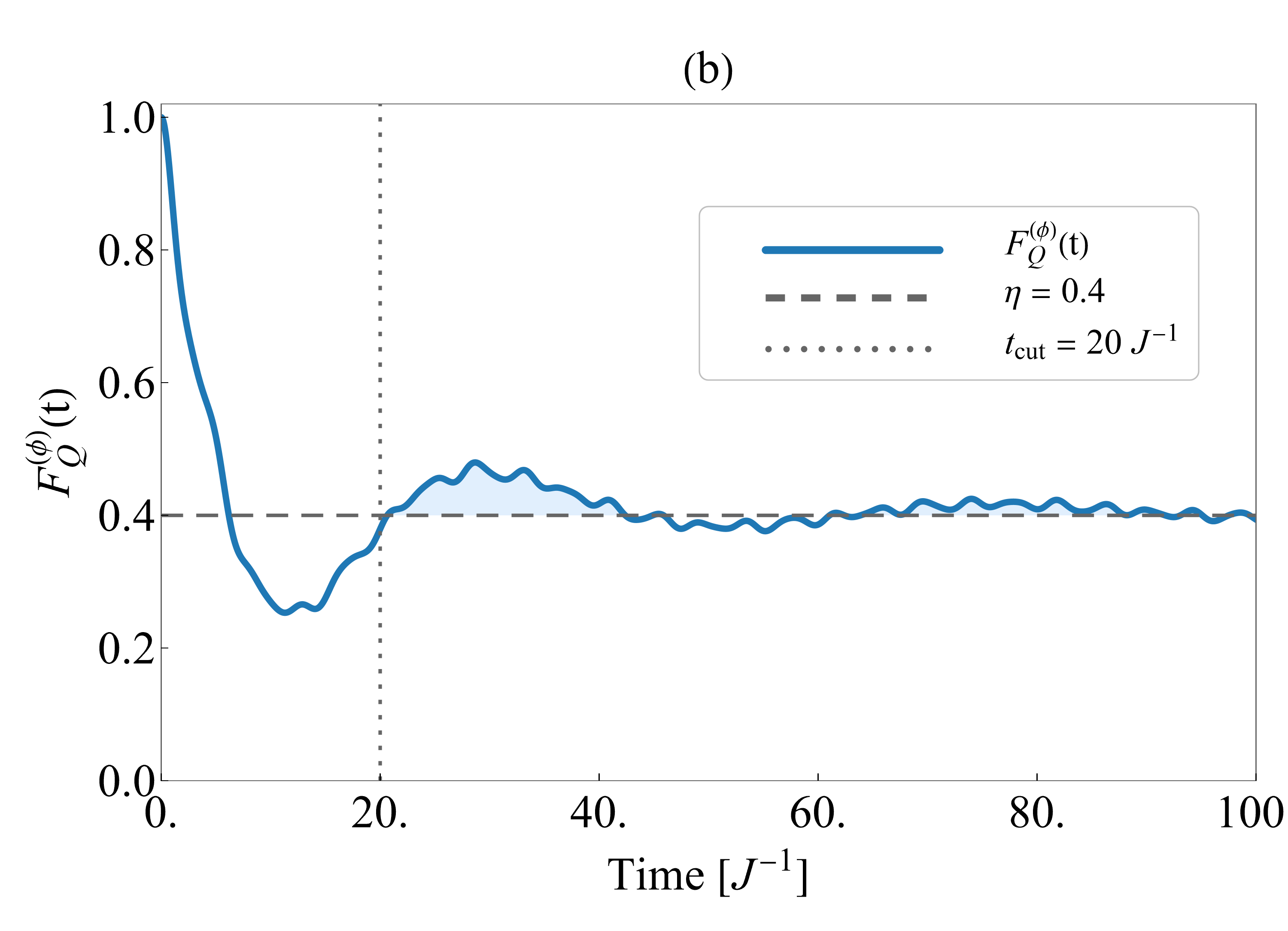}\\
    \includegraphics[width=\columnwidth]{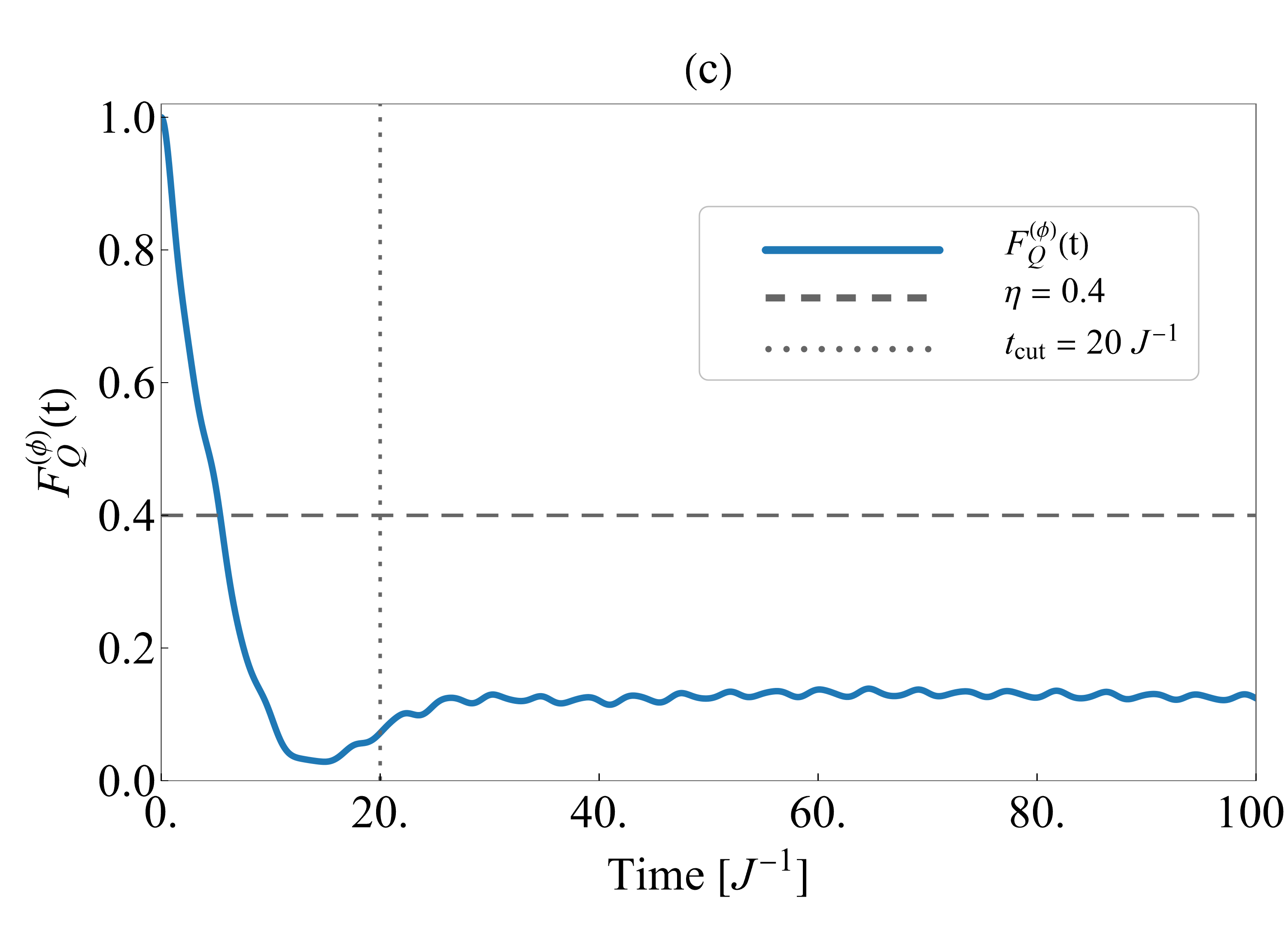}
    \caption{Time-dependent phase-QFI $F_Q^{(\phi)}(t)$ for three representative detunings at fixed $d=0.3$ and $g=0.4J$. The shaded regions indicate the portions of the dynamics for which $F_Q^{(\phi)}(t)\ge \eta$ after the initial transient. The useful-window diagnostic is defined by Eq.~\eqref{eq:useful_window} with threshold $\eta=0.4$, cutoff $t_{\mathrm{cut}}=20J^{-1}$, and observation horizon $T=100J^{-1}$. Panels (a)--(c) correspond to (a) a deep in-gap frequency, $\Delta/(2J|d|)=0$, for which $W_{0.4}=80\,J^{-1}$, (b) a near-edge configuration, $\Delta/(2J|d|)=0.95$, for which $W_{0.4}\simeq 55.9\,J^{-1}$, and (c) a continuum configuration, $\Delta/(2J|d|)=1.20$, for which $W_{0.4}=0$. Numerical results are obtained for a finite SSH chain with $2L+1$ unit cells, $L=500$, and a Krylov subspace of dimension $350$.}
    \label{fig:useful_window}
\end{figure}

Figure~\ref{fig:useful_window} applies this diagnostic to three representative detunings at fixed $d=0.3$ and $g=0.4J$, using the same observation horizon $T=100J^{-1}$ employed throughout the manuscript, a post-transient cutoff $t_{\mathrm{cut}}=20J^{-1}$, and a threshold $\eta=0.4$. The three panels summarize the gap-to-continuum crossover in a form that is directly useful for experiment design. Deep in the gap ($\Delta=0$, i.e.\ $\Delta/(2J|d|)=0$), the phase-QFI rapidly settles into a wide nondecaying window, so that essentially the entire post-transient interval remains metrologically usable and $W_{0.4}=80\,J^{-1}$. Near the band edge ($\Delta/(2J|d|)=0.95$, i.e.\ $\Delta/J=0.57$), the signal still recovers after the initial dip but the useful window narrows to $W_{0.4}\simeq 55.9\,J^{-1}$. In the continuum ($\Delta/(2J|d|)=1.20$, i.e.\ $\Delta/J=0.72$), by contrast, the post-transient signal never re-enters the target band and the useful window collapses to $W_{0.4}=0$.

This construction is complementary to both $t_\eta$ and $\overline{F}_Q^{(\phi)}$: the retention time emphasizes the first loss of a chosen threshold, the late-time average summarizes the overall post-transient magnitude, and $W_\eta$ measures how much scheduling freedom the experiment retains once the initial transient has passed. In the SSH bath, this freedom is large only when the microscopic spectrum supports a sufficiently strong nondecaying component. The same bound-state support that sustains the retained late-time signal therefore also opens a broad interrogation window robust to timing uncertainty.

\subsection{Detuning-resolved bound-state diagnostics}
\label{sec:results_boundstate}

The off-resonant analysis above showed that the late-time metrological response is organized by the position of the emitter relative to the SSH gap edge, while Sec.~\ref{sec:analytic_offresonant} identified the central in-gap bound-state residue as the main analytical benchmark for the retained signal. It is therefore natural to ask how closely the finite-window late-time indicators used in Sec.~\ref{sec:results_offresonant} follow this bound-state prediction, and how the underlying in-gap state reorganizes as the detuning is varied. To answer this, we extract from the finite-chain spectrum, for each $\Delta$, the total emitter residue carried by the in-gap eigenstate,
\begin{equation}
Z_{\mathrm{BS}}(\Delta)=\sum_{\alpha\in \mathrm{gap}} |\langle e|\psi_\alpha\rangle|^2,
\label{eq:zbs_detuning}
\end{equation}
and the corresponding separation from the nearest band edge,
\begin{equation}
\delta_{\mathrm{edge}}(\Delta)=2J|d|-|E_{\mathrm{BS}}(\Delta)|,
\label{eq:delta_edge}
\end{equation}
where $E_{\mathrm{BS}}$ is the in-gap bound-state energy with the largest emitter overlap. These two quantities resolve, respectively, how much of the emitter is carried by the nondecaying component and how deeply that component lies inside the SSH gap.

\begin{figure}[tbp]
    \centering
    \includegraphics[width=\columnwidth]{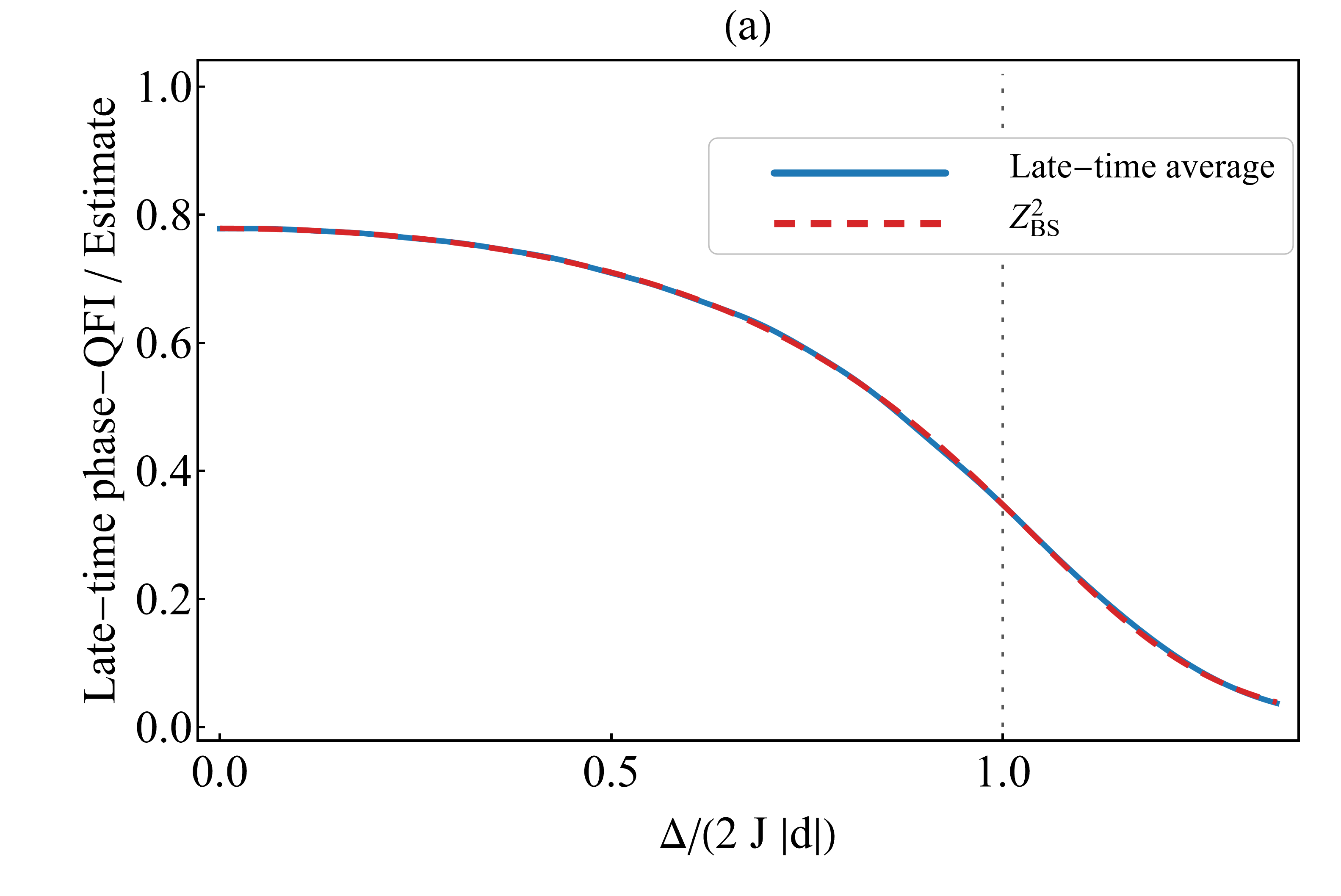}\\
    \includegraphics[width=\columnwidth]{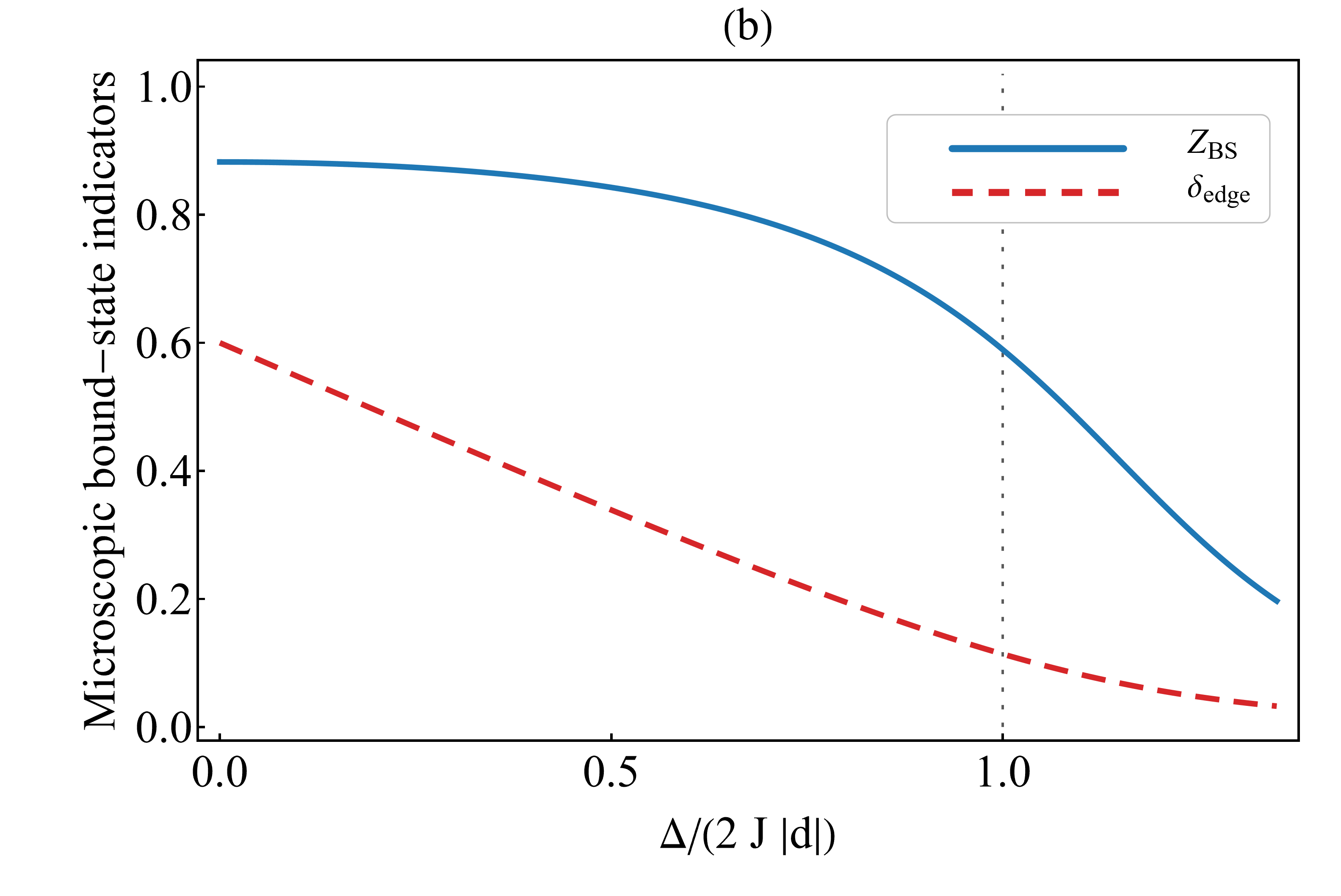}
    \caption{Detuning-resolved bound-state diagnostics for $d=0.3$ and $g=0.4J$. (a) Late-time averaged phase-QFI $\overline{F}_Q^{(\phi)}$ over the window $40J^{-1}\le t\le 100J^{-1}$ together with the central in-gap bound-state estimate $Z_{\mathrm{BS}}^2$. (b) Emitter residue $Z_{\mathrm{BS}}$ and distance $\delta_{\mathrm{edge}}$ of the in-gap state from the nearest SSH band edge, as defined in Eq.~\eqref{eq:delta_edge}. The vertical dotted line marks the bare condition $\Delta/(2J|d|)=1$. Numerical diagonalizations use a finite SSH chain with $2L+1$ unit cells, $L=220$, via exact diagonalization of the single-excitation Hamiltonian.}
    \label{fig:detuning_boundstate}
\end{figure}

Figure~\ref{fig:detuning_boundstate} shows the outcome for $d=0.3$ and $g=0.4J$. Panel~(a) compares the late-time averaged phase-QFI with the central in-gap bound-state estimate $Z_{\mathrm{BS}}^2$ across a broad detuning sweep. The two curves track one another closely throughout the scan, including beyond the bare band-edge point $\Delta/(2J|d|)=1$, where the in-gap state still exists but becomes progressively more weakly localized on the emitter. Panel~(b) shows the corresponding microscopic trend: both the emitter residue $Z_{\mathrm{BS}}$ and the band-edge distance $\delta_{\mathrm{edge}}$ decrease monotonically as the emitter is detuned away from the gap center. The loss of metrological protection is therefore not merely a kinematic consequence of crossing the nominal gap edge. It reflects the gradual erosion of the emitter-supported in-gap state predicted analytically by Eqs.~\eqref{eq:pole_equation}--\eqref{eq:ZBS_residue_general}.

This detuning-resolved comparison connects the Green's-function analysis with the operational diagnostics introduced above. Equation~\eqref{eq:FQ_ZBS_exact} gives the central in-gap bound-state contribution, while Fig.~\ref{fig:detuning_boundstate}(a) shows that the finite-window indicator $\overline{F}_Q^{(\phi)}$ follows essentially the same detuning dependence across the numerically accessible regime. Panel~(b) then identifies the underlying microscopic mechanism explicitly: as $\Delta$ is displaced from the gap center, the in-gap bound state moves toward the nearest band edge, its emitter residue decreases, and the late-time metrological protection degrades accordingly. In this sense, the SSH bath does more than provide a generic spectral gap: it offers a microscopic lattice realization in which the late-time phase sensitivity, the available interrogation time, and the detuning dependence of the bound-state residue are all organized by the same spectral mechanism.

The overall picture is therefore consistent across all observables considered here. Dimerization sets the gap scale, detuning fixes the emitter position relative to the gap and the band edges, and the resulting bound-state content controls the late-time metrological protection. The SSH bath thus provides a microscopic lattice realization in which late-time phase sensitivity, operational interrogation windows, and detuning-dependent bound-state support are organized by the same spectral mechanism.

\section{Conclusions}
\label{sec:conclusions}

We have studied phase metrology for a single quantum emitter coupled to a bosonic SSH environment and developed a microscopic account of how a structured lattice spectrum governs the retention of phase information. Across the parameter range considered here, the metrological response is controlled by four linked ingredients: the width of the SSH central gap, the spectral position of the emitter relative to the gap and band edges, the emitter-bath coupling strength, and the bound-state weight supported by the coupled emitter-bath system.

At resonance, the SSH gap sustains a finite late-time phase-QFI that can be obtained in closed form and tuned through the dimerization-controlled gap scale. The same resonant analysis shows that the emitter-bath coupling strength plays a distinct role: at fixed dimerization, increasing $g$ enhances transient hybridization but reduces the retained late-time signal by lowering the emitter weight carried by the in-gap bound state. More generally, a Dyson-equation analysis of the local bath Green's function and the in-gap pole equation identifies the central in-gap bound state and its emitter residue at arbitrary detuning. This residue provides the main bound-state benchmark for the post-transient phase-QFI in the regimes considered here. It also explains the sign independence of the protocol through the $d^2$ structure of the local Green's function and predicts the suppression of the retained signal as the bound state approaches a band edge and its emitter residue vanishes.

Away from resonance, late-time averages, retention times, and post-transient useful-window diagnostics resolve a crossover between a gap-protected regime, an edge-sensitive crossover region, and a continuum regime in which late-time metrological protection is strongly reduced. The comparison with the uniform-chain limit shows that the retained resonant signal disappears when the SSH gap closes, while the detuning-resolved bound-state analysis shows that the finite-window late-time signal closely follows the detuning dependence of the central in-gap residue. Together, these results show that the SSH bath does more than provide a generic spectral gap: it offers a microscopic lattice setting in which late-time phase sensitivity, operational interrogation windows, and detuning-dependent bound-state support are organized by the same spectral mechanism.

The role of the SSH setting lies in its microscopic character: dimerization, detuning, and local emitter coupling all enter explicitly at the Hamiltonian level, so the protection of phase information is tied to a lattice model rather than to an externally imposed spectral function. At the same time, the bulk-sign-symmetry analysis makes clear the scope of the present protocol: in the local single-emitter geometry studied here it does not resolve the SSH winding sector. We note that metrological signatures that are explicitly sensitive to the winding sector would require modified geometries. For example, the use of boundary-coupled or multi-emitter configurations in which the coupling is sensitive to the relative phase between the two sublattices~\cite{TopoSensor2025,Vega2023}. This is beyond the scope of the present bulk, local, single-emitter analysis.

\bibliography{references_bib_audited}

\end{document}